\documentclass[aap,MSNbibl,citesort,dvips]{arximspdf}
\usepackage{dcolumn}
\doi{10.1214/09-AAP636}
\volume{20}
\issue{3}
\pubyear{2010}
\firstpage{841}
\lastpage{868}

\makeatletter
\DeclareMathAlphabet\mathcaligr{OMS}{cmsy}{m}{n}
\newcolumntype{d}[1]{D{.}{.}{#1}}

\newcommand{\eps}{\varepsilon}
\newcommand{\To}{\longrightarrow}
\newcommand{\B}{\mathcaligr{B}}
\newcommand{\M}{\mathcaligr{M}}
\newcommand{\p}{\mathbb{P}}
\newcommand{\E}{\mathbb{E}}
\newcommand{\X}{\mathcaligr{X}}
\newcommand{\F}{\mathcaligr{F}}
\newcommand{\Real}{\mathbb{R}}
\newcommand{\abs}[1]{ \vert #1 \vert}
\newcommand{\abss}[1]{ \big\vert #1 \big\vert}
\newcommand{\nnnorm}[1]{\big\vert\!\big\vert\!\big\vert#1\big\vert\!\big\vert\!\big\vert}
\newcommand{\nnorm}[1]{\vert\!\vert\!\vert#1\vert\!\vert\!\vert}
\newcommand{\C}{\mathcaligr{C}}
\newcommand{\D}{\mathcaligr{D}}

\newcommand{\PP}{\mathbb{P}} % proba
\newcommand{\PE}{\mathbb{E}} % esperance

\newtheorem{theorem}{Theorem}[section]
\newtheorem{corollary}{Corollary}[section]
\newtheorem{proposition}{Proposition}[section]

\newtheorem{lemma}{Lemma}[section]
\newtheorem{assumption}{Assumption}[section]
\newproclaim{remark}{Remark}[section]
\newproclaim{algo}{Algorithm}[section]
\makeatother

\begin{document}
\begin{frontmatter}

\title{A cautionary tale on the efficiency of some adaptive Monte
Carlo schemes}
\runtitle{On the Efficiency of some adaptive Monte Carlo schemes}

\begin{aug}
\author[A]{\fnms{Yves F.} \snm{Atchad\'{e}}\ead[label=e1]{yvesa@umich.edu}}
\runauthor{Y. F. Atchad\'{e}}
\pdfauthor{Yves F. Atchade}
\affiliation{University of Michigan}
\address[A]{Department of Statistics\\
University of Michigan\\
1085 South University\\
Ann Arbor, Michigan 48109\\
USA\\
\printead{e1}} %adresu isvedimo komanda gale!
\end{aug}

% HISTORY:
\received{\smonth{9} \syear{2007}}
\revised{\smonth{8} \syear{2009}}

% ABSTRACT
%
\begin{abstract}
There is a growing interest in the literature for adaptive Markov chain
Monte Carlo methods
based on sequences of random transition kernels $\{P_n\}$ where the
kernel $P_n$ is allowed
to have an invariant distribution $\pi_n$ not necessarily equal to the
distribution of
interest $\pi$ (target distribution). These algorithms are designed
such that as $n\to\infty$,
$P_n$ converges to $P$, a kernel that has the correct invariant
distribution $\pi$. Typically,
$P$ is a kernel with good convergence properties, but one that cannot
be directly implemented.
It is then expected that the algorithm will inherit the good
convergence properties of $P$.
The equi-energy sampler of [\textit{Ann. Statist.} \textbf{34} (2006)
1581--1619] is an example of this type of adaptive MCMC.
We show in this paper that the asymptotic variance of this type of
adaptive MCMC is always
at least as large as the asymptotic variance of the Markov chain with
transition kernel $P$.
We also show by simulation that the difference can be substantial.
\end{abstract}

% KEYWORDS
%

\end{frontmatter}

%s1 ###
\section{Introduction}\label{intro}
Adaptive Markov chain Monte Carlo (AMCMC) is an approach to Markov
chain Monte Carlo (MCMC) simulation where the transition kernel of the
algorithm is allowed to change over time as an attempt to improve
efficiency. It grows out of the seminal works of \cite
{gilksetal98,haarioetal00}. Let $\pi$ be the distribution of interest.
The problem is to sample efficiently from $\pi$ given a family of
Markov kernels $\{P_\theta,\theta\in\Theta\}$. This can be solved
adaptively using a joint process $\{(X_n,\theta_n),n\geq0\}$ such
that the conditional distribution of $X_{n+1}$ given the information
available up to time $n$ is $P_{\theta_n}$ and where $\theta_n$ is
adaptively tuned over time. Some general sufficient conditions for the
convergence of such algorithms can be found in \cite
{rosenthaletroberts05,atchadeetfort08}. It is also shown in \cite
{andrieuetatchade05} that under some regularity conditions, if a
``best'' limiting kernel $P_{\theta^*}$ exists, the marginal chain $\{
X_n,n\geq0\}$ in the joint adaptive process behaves in many ways like
a standard Markov chain with transition kernel $P_{\theta^*}$. In all
the above-mentioned papers, the assumption that each $P_\theta$ has
invariant distribution $\pi$ plays an important role.

More recently, interest has emerged in building Monte Carlo algorithms
where the transition kernel $P_n$ used at time $n$ has invariant
distribution $\pi_n$ not necessarily equal to $\pi$. These algorithms
are designed such that as $n\to\infty$, $P_n$ converges to a
transition kernel $P$ which is invariant with respect to $\pi$. This
limiting kernel $P$ is typically a very efficient kernel that would be
difficult to implement otherwise. The interest of this approach is that
as $n\to\infty$, $P_n$ approaches $P$ and one expects the algorithm
to inherit the good convergence properties of $P$. The equi-energy (EE)
sampler of \cite{kzw06} is an example. Another example based on
importance resampling appeared independently in \cite{andrieuetal07} and
\cite{atchade06}.

This paper provides a detailed analysis of the law of large numbers and
central limit theorem for the EE sampler. It is also an attempt to
address the question of whether such algorithms can deliver the same
performance as their limiting kernel~$P$.
%This algorithm uses the idea of \textit{tempering} (
%of the target distribution to form empirical measures approximations
%of that target distribution.
%The idea is intuitively appealing and bears some similarity with other
%tempering algorithms like \textit{parallel tempering} (
We give a negative answer. We show, in the case of the EE sampler, that
its asymptotic variance is always at least as large as the asymptotic
variance of the limiting transition kernel $P$. The difference can be
substantial and we illustrate this with a simulation example.

On the related literature, the law of large numbers for of the EE
sampler has been studied in \cite{andrieuetal07} but using different
techniques than those in this work. We also mention a new class of
interacting MCMC algorithms proposed by \cite
{doucetetdelmoral08a,doucetetdelmoral08b} for solving numerically some
discrete-time measure-valued equations. These algorithms share the same
framework with the EE sampler. In these two papers, the authors develop
a number of asymptotic results for interacting MCMC including a strong
law of large numbers and a central limit theorem.
%The asymptotic variances for these interacting MCMC also show, as in
%the case of the EE sampler developed here, that there is a cost in
%learning the properties of the limiting distribution.

The paper is organized as follows. In Section \ref{secalgo} we present
the EE sampler and IR-MCMC in a slightly more general framework. The
limit theorems are developed in Section \ref{limittheory} and proved
in Section \ref{proof}. The main ingredient of the proofs is the
martingale approximation method. We present a simulation example in
Section \ref{ex1} comparing these algorithms to a Random Walk
Metropolis algorithm.

\section{A class of adaptive Monte Carlo algorithms}\label{secalgo}

Let $(\X,\B,\lambda)$ be a reference Polish space equipped with its
Borel $\sigma$-algebra $\B$ and a $\sigma$-finite measure $\lambda$
and $K\geq1$ an integer. We denote by $\M$ the set of all probability
measure on $(\X,\B)$. Let $\{\pi^{(l)},l=0,\ldots,K\}$ be
probability measures on $(\X,\B)$ such that
%e1 ###
%
\begin{equation}\label{pil}
\pi^{(l)}(dx)=\frac{1}{Z_l}e^{-E_l(x)}\lambda(dx)
\end{equation}
for some measurable functions $E_l\dvtx(\X,\B)\to\Real$.
$Z_l:=\int_\X e^{-E_l(x)}\lambda(dx)$ (assumed finite) is the
normalizing constant. We study a class of Monte Carlo algorithms to
sample from the family $\{\pi^{(l)}\}$. These algorithms will
generated an ergodic random process $\{(X_n^{(0)},\ldots
,X_n^{(K)}),n\geq0\}$ on $\X^{K+1}$ with limiting distribution $\pi
^{(0)}\times\cdots\times\pi^{(K)}$.

We introduce some notation in order to describe the algorithm.
Whenever necessary and without further notice, any subset of $\Real^d$ will
be equipped with its Borel $\sigma$-algebra. If $(\mathcaligr
{Y},\mathcaligr{E})$ and
$(\mathcaligr{Z},\F)$ are two measurable spaces, a kernel from
$(\mathcaligr{Y},\mathcaligr{E})$
to $(\mathcaligr{Z},\F)$ is any function $P\dvtx \mathcaligr
{Y}\times\mathcaligr{F}\to[0,1]$
such that $P(y,\cdot)$ is a probability measure on $(\mathcaligr
{Z},\F)$ for all
$y\in\mathcaligr{Y}$ and $P(\cdot,A)$ is a measurable map for all
$A\in\mathcaligr{F}$.
If $(\mathcaligr{Y},\mathcaligr{E})=(\mathcaligr{Z},\F)$, we call
$P$ a kernel on
$(\mathcaligr{Z},\F)$. If $P$ is a kernel from $(\mathcaligr
{Y},\mathcaligr{E})$ to
$(\mathcaligr{Z},\F)$, $f\dvtx (\mathcaligr{Z},\F)\to\Real$ a
measurable function and
$y\in\mathcaligr{Y}$, we shall use the notation $P(y,f)$ or $Pf(y)$
to denote the integral
$\int_{\mathcaligr{Z}}P(y,dz)f(z)$ whenever it is well defined.

%s2.1 ###
\subsection{A general algorithm}
Let $\{P^{(l)}, l=0,\ldots,K\}$ be kernels on $(\X,\B)$ such that
$\pi^{(l)}$ is the invariant distribution of $P^{(l)}$. Let $\{
T^{(l)}, l=1,\ldots,K\}$ be kernels from $(\X^2,\B^2)$ to $(\X,\B
)$, $\{\omega^{(l)}, l=1,\ldots,K\}$ positive real-valued measurable
functions defined on $(\X^2,\B^2)$ and $\theta_l\in(0,1)$ for
$l=1,\ldots,K$.
For $\mu\in\M$ and $l=1,\ldots,K$, we define the following kernel
on $(\X,\B)$:
%e2 ###
%
\begin{eqnarray}\label{Pmu}\hspace*{20pt}
P^{(l)}_\mu(x,A)&=&\theta_lP^{(l)}(x,A)\nonumber\\[-8pt]\\[-8pt]\hspace*{20pt}
&&{}+(1-\theta_l)\frac{\int\mu
(dy)\omega^{(l)}
(y,x)T^{(l)}(y,x,A)}{\int\mu(dy)\omega^{(l)}(y,x)},\qquad x\in\X
,A\in\B.\nonumber
\end{eqnarray}
For $n\geq1$, we introduce the maps $H_n\dvtx\M\times\X\to\M$ defined
as $H_n(\mu,x)=\mu+n^{-1}(\delta_x-\mu)$, where $\delta_x$ is the Dirac
measure. Let $\{(X_n^{(0)},\ldots,X_n^{(K)},\mu^{(0)}_n,\ldots,\break\mu
_n^{(K-1)}), n\geq0\}$
be the nonhomogeneous Markov chain on $\X^{K+1}\times\M^{K}$
[defined on some probability space $(\Omega,\F)$ that can be taken as
the canonical space $(\X^{K+1}\times\M^{K})^\infty$] with sequence
of transition kernels $\bar P_n$ given by
\begin{eqnarray}
&&\bar P_n \bigl( \bigl(x^{(0)},\ldots,x^{(K)},\mu^{(0)},\ldots,
\mu^{(K-1)} \bigr);\nonumber\\
&&\hspace*{16pt}\bigl(dy^{(0)},\ldots,dy^{(K)},d\nu
^{(0)},\ldots,d\nu^{(K-1)} \bigr) \bigr)\\
&&\qquad=P^{(0)}\bigl(x^{(0)},dy^{(0)}\bigr)\prod
_{l=1}^KP^{(l)}_{\mu^{(l-1)}} \bigl(x^{(l)},dy^{(l)}
\bigr)\prod_{l=0}^{K-1}\delta_{H_n(\mu^{(l)},y^{(l)})}\bigl(d\nu
^{(l)}\bigr).\nonumber
\end{eqnarray}
Throughout, we denote $\{\F_n, n\geq0\}$ the natural filtration of
the process. We will assume that the initial value of the process is
fixed. For simplicity we take $\mu_0^{(l)}=0$. Finally, we call $\PP$
and $\PE$ the probability distribution and expectation of the process.

Algorithmically, $\{(X_n^{(0)},\ldots,X_n^{(K)},\mu^{(0)}_n,\ldots
,\mu_n^{(K-1)}), n\geq0\}$ can be described as follows.

\begin{algo}\label{algo}
At time $n$ and given $\{(X_k^{(0)},\ldots,X_k^{(K)},\mu
_k^{(0)},\ldots,\mu_k^{(K-1)}),\break k\leq n-1\}$:
\begin{enumerate}
\item Generate $X_{n}^{(0)}\sim P^{(0)}(X_{n-1}^{(0)},\cdot)$.
\item For $l=1,\ldots,K$, generate independently $X_{n}^{(l)}$ from
$P_{\mu_{n-1}^{(l-1)}}^{(l)}
(X_{n-1}^{(l)},\cdot)$ as given by~(\ref{Pmu}).
\item For $l=0,\ldots,K-1$, set $\mu_n^{(l)}=H_n (\mu
_{n-1}^{(l)},X_n^{(l)} )
=\mu_{n-1}^{(l)}+n^{-1} (\delta_{X_n^{(l)}}-\mu_{n-1}^{(l)} ).$
\end{enumerate}
\end{algo}

The heuristic of the algorithm is the following. By construction, $\{
X^{(0)}_n,\F_n\}$ is a Markov chain with kernel $P^{(0)}$ and
invariant distribution $\pi^{(0)}$. If this chain is ergodic, then as
$n\to\infty$, $\PP(X_n^{(1)}\in A|\F_{n-1} )=P^{(1)}_{\mu
_{n-1}^{(l-1)}}(X_{n-1}^{(l)},A)$, will\vspace*{-4pt} converge to $K^{(1)}$ where
$K^{(l)}$ is given by
%e3 ###
%
\begin{eqnarray}\label{Plim}
K^{(l)}(x,A)&=&\theta_lP^{(l)}(x,A)\nonumber\\[-8pt]\\[-8pt]
&&{}+
(1-\theta_l)\frac{1}{z^{(l)}(x)}\int_{\X}\pi^{(l-1)}(dy)\omega
^{(l)}(x,y)T^{(l)}(y,x,A),\nonumber
\end{eqnarray}
where $z^{(l)}(x)=\int_{\X}\pi^{(l-1)}(dy)\omega^{(l)}(x,y)$. We
will discuss\vspace*{1pt} below two ways of choosing $\omega^{(l)}$ and $T^{(l)}$
so that $K^{(l)}$ has invariant distribution $\pi^{(l)}$. With these
choices we can reasonably expect $\{X^{(1)}_n\}$ to be ergodic with
limiting distribution $\pi^{(1)}$. The same argument can then be
repeated. In other words, with appropriate choice of $\omega^{(l)}$
and $T^{(l)}$, the marginal process $\{X_n^{(l)}, n\geq0\}$ can be
used for Monte Carlo simulation from $\pi^{(l)}$.

%The choice $\omega^{(l)}(x,y)=r^{(l)}(y)$ and
%$T^{(l)}(y,x,A)=T^{(0)}(y,A)$, where $T^{(0)}$ is a transition kernel
%on $(\X,\)$ with invariant distribution $\pi^{(l)}$ gives the IR-MCMC,
%whereas The interest of the approach is that the limiting kernel
%$K^{(l)}$ can have strickingly better convergence properties than
%$P^{(l)}$ (at least in the case of the EE sampler and IR-MCMC) but is
%typically impossible to implement directly. This naturally raises the
%question of interest here, does $\{X_n^{(l)}\}$ behaves asymptotically
%like a Markov chain with kernel $K^{(l)}$? Before considering the
%question, we describe the EE sampler and IR-MCMC in more detail.

%s2.2 ###
\subsection{Importance-resampling MCMC}\label{irmcmc}
For $l=1,\ldots,K$ define the importance function
\[
r^{(l)}(x)=\exp\bigl(E_{l-1}(x)-E_{l}(x) \bigr).
\]
In Algorithm \ref{algo} we can take $\omega^{(l)}(x,y)=r^{(l)}(y)$ and
$T^{(l)}(y,x,A)=T_0^{(l)}(y,A)$, where $T_0^{(l)}$ is some kernel on
$(\X,\B)$
with invariant distribution $\pi^{(l)}$. This leads to the IR-MCMC algorithm
(\cite{atchade06,andrieuetal07}). In this case, step~2 of
Algorithm~\ref{algo} can be described as follows: with probability
$\theta_l$ we sample $X^{(l)}_{n}$ from $P^{(l)}(X^{(l)}_{n-1},\cdot
)$ and with probability $1-\theta_l$, we obtain $Y^{(l)}$ by
resampling from $\{X_0^{(l-1)},\ldots,X_{n-1}^{(l-1)}\}$
with weights $\{r^{(l)}(X_0^{(l-1)}),\ldots,r^{(l)}(X_{n-1}^{(l-1)})\}
$ and then propose $X_{n}^{(l)}\sim T_0^{(l)}(Y^{(l)},\cdot)$.

The $l$th limiting kernel here takes the form
\[
K^{(l)}(x,A)=\theta_lP^{(l)}(x,A)+(1-\theta_l)\pi^{(l)}(A)
\]
has invariant distribution $\pi^{(l)}$ and has better mixing than
$P^{(l)}$. But direct sampling from $K^{(l)}$ is impossible as it
requires that we be able to sample from $\pi^{(l)}$ which is the
problem that we are trying to solve in the first place.

%s2.3 ###
\subsection{The EE sampler}\label{Sectionee}

Taking $\omega^{(l)}(x,y)\equiv1$ and
%e4 ###
%
\begin{equation}\label{Tlee}
T^{(l)}(y,x,A)=\min\biggl(1,\frac{r_l(y)}{r_l(x)} \biggr)\mathbf
{1}_A(y)+ \biggl(1-\min\biggl(1,\frac{r_l(y)}{r_l(x)} \biggr) \biggr)\mathbf{1}_A(x),
\end{equation}
in (\ref{Pmu}), we get the EE sampler (\cite{kzw06}). In this case
the limiting kernel becomes
%e5 ###
%
\begin{eqnarray}\label{Plimee}
K^{(l)}(x,A)
&=&\theta_lP^{(l)}(x,A)+(1-\theta_l)\int_{\X}\pi
^{(l-1)}(dy)T^{(l)}(y,x,A)\nonumber\\[-8pt]\\[-8pt]
&=&\theta_lP^{(l)}(x,A)+(1-\theta_l)R^{(l)}(x,A),\nonumber
\end{eqnarray}
where $R^{(l)}$ is the kernel of the Metropolis--Hastings algorithm
with proposal $\pi^{(l-1)}$ and target distribution $\pi^{(l)}$:
\begin{eqnarray*}
R^{(l)}(x,A)&=&\int_A\min\biggl(1,\frac{r^{(l)}(y)}{r^{(l)}(x)}
\biggr)\pi^{(l-1)}(dy)\\
&&{}+
\biggl[1-\int_\X\min\biggl(1,\frac{r^{(l)}(y)}{r^{(l)}(x)} \biggr
)\pi^{(l-1)}(dy) \biggr]
\mathbf{1}_A(x).
\end{eqnarray*}
Clearly, $K^{(l)}$ has invariant distribution $\pi^{(l)}$. In general
$K^{(l)}$ will converge faster than $P^{(l)}$. For example if
$E_l-E_{l-1}$ is bounded from below it is easy to show that $K^{(l)}$
is always uniformly ergodic, independently of $P^{(l)}$.

For the EE sampler, step~2 of Algorithm \ref{algo} can now be
described as follows. With probability $\theta_l$ we sample
$X^{(l)}_{n}$ from $P^{(l)}(X^{(l)}_{n-1},\cdot)$ and with probability
$1-\theta_l$,
we obtain $Y^{(l)}$ by resampling uniformly from $\{X_k^{(l-1)}\dvtx
k\leq n-1\}$. Then $Y^{(l)}$ is accepted with probability $\min
(1,\frac{r^{(l)}(Y^{(l)})}{r^{(l)}(X^{(l)}_{n-1})} )$ in which case we
set $X_{n}^{(l)}=Y^{(l)}$; otherwise $Y^{(l)}$ is rejected and we set
$X_{n}^{(l)}=X^{(l)}_{n-1}$.

Actually the EE sampler described above is a simplified version of
\cite{kzw06}.
Their original algorithm uses an idea of partitioning. Let $\{\X_i,
i=1,\ldots,d\}$
be a partition of $\X$ (in \cite{kzw06}, $E_l(x)=E(x)/t_l$ and they
take $\X_i=\{x\in\X\dvtx
E_{i-1}<E(x)\leq E_{i}\}$ for some predefined valuse $E_0<E_1<\cdots<
E_{d}$). Define the function
$I(x)=i$ if $x\in\X_i$; so $\X_{I(x)}$ represents the component of
the partition
to which $x$ belongs. Now set $\omega^{(l)}(x,y)=\mathbf{1}_{\X
_{I(x)}}(y)$ and $T^{(l)}$ as
in (\ref{Tlee}) and we get the EE sampler of \cite{kzw06}. In this
general case, the limiting
kernel has the same form as in (\ref{Plimee}) but where $R^{(l)}$ is
now a Metropolis--Hastings algorithm with target distribution\vspace*{1pt} $\pi
^{(l)}$ and proposal kernel $Q^{(l)}(x,dy)\propto\pi
^{(l-1)}(y)\mathbf{1}_{\X_{I(x)}}(y)\lambda(dy)$. Partitioning the
state\vspace*{1pt} space and using the proposal $Q^{(l)}(x,dy)\propto\pi
^{(l-1)}(y)\mathbf{1}_{\X_{I(x)}}(y)\lambda(dy)$ works well in
practice as it can allow large jumps in the state space to be accepted.
But it does not add any significant feature to the algorithm from the
theoretical standpoint. Therefore and to simplify the analysis, we only
consider the case where no partitioning is used ($\X_{I(x)}=\X$ for
all $x\in\X$).

%s3 ###
\section{Asymptotics of the EE sampler}\label{limittheory}

For the remaining of the paper, we restrict our attention to the EE sampler.
In other words, we consider the process defined in Section \ref
{secalgo} with
$\omega^{(l)}(x,y)\equiv1$ and $T^{(l)}$ as defined in (\ref{Tlee}).

%s3.1 ###
\subsection{Notation and assumptions}

We start with some notation. If $P_1,P_2$ are kernels on $(\X,\B)$,
the product $P_1P_2$ is the kernel $P_1P_2(x,A)=\break\int_\X
P_1(x,dy)P_2(y, A)$. If $\mu$ is a signed measure on $(\X,\B)$, we
write $\mu(f)$ to denote the integral $\int\mu(dx)f(x)$ and we will
also use $\mu$ to denote the linear functional on the space of $\Real
$-valued functions on $(\X,\B)$ thus induced.
Similarly, we will write $\mu P_1(A)$ for $\int\mu(dx)P_1(x,A)$. Let
$V\dvtx\X\to[1,\infty)$ be given. For $f\dvtx(\X,\B)\to\Real$,
we define its $V$-norm as $\abs{f}_V:=\sup_{x\in\X}\frac{\abs
{f(x)}}{V(x)}$ and we introduce the space $L^\infty_{V}$ of measurable
real-valued functions defined on $\X$ such that $\abs{f}_V<\infty$.
For a signed measure $\mu$ on $(\X,\B)$ we define by $\Vert\mu
\Vert_V:=\sup\{\abs{\mu(f)}, f\in L^\infty_V, \abs{f}_V\leq1\}$. We
equip $\M$, the set of all probability measures on $(\X,\B)$, with
the metric $\Vert\mu-\nu\Vert_V$ and the Borel $\sigma$-algebra $\B
_{\M}(V)$ induced by $\Vert\cdot\Vert_V$. Whenever $V$ is understood,
we will write $(\M,\B_\M)$ instead of $(\M,\B_{\M}(V))$.
For a linear operator $T$ from $(L^\infty_V,\abs{\cdot}_V)$ into
itself, we define its operator norm by $\nnorm{T}_V:=\sup\{\abs
{Tf}_V, f\in L^\infty_V, \abs{f}_V\leq1\}$.

%We assume that $\X$ is a Polish space (a complete separable metric
%space) and that $\B$ is its Borel $\sigma$-algebra.
We assume that $\pi^{(l)}$ is of the form
%e6 ###
%
\begin{equation}\pi^{(l)}(dx)=\frac{1}{Z_l}e^{-E(x)/t_l}\lambda(dx)
\end{equation}
for some continuous function $E\dvtx(\X,\B)\to\Real$ that is
bounded from below and $t_1>\cdots>t_K=1$ is a decreasing sequence of
positive numbers (temperatures). In addition, we make the following assumption.
\renewcommand{\theassumption}{(A\arabic{assumption})}
\begin{assumption}\label{assum1}
For $l=1,\ldots,K$, there exist a set $C_l \subset\X$,
a probability measure $\phi_l$ such that $\phi_l(C_l)>0$ an integer
$n_0>0$ and constants
$\lambda_l \in(0,1)$, $b_l\in[0,\infty)$, $\eps_l \in(0,1]$ such
that for $x \in\X$
and $A\in\B$,
%e7 ###
%
\begin{equation}\label{mino}
\bigl[P^{(l)} \bigr]^{n_0}(x,A)\geq\eps_l\phi_l(A)\mathbf{1}_{C_l}(x)
\end{equation}
and
%e8 ###
%
\begin{equation}\label{drift}
P^{(l)}V(x)\leq\lambda_l V(x)+b_l\mathbf{1}_{C_l}(x) ,
\end{equation}
where $V(x)=ce^{\kappa E(x)}\geq1$ for some finite constants $c>0$ and
$\kappa\in(0,1)$ and $0<\kappa< (\frac{1}{t_l}-\frac{1}{t_{l-1}}
)$. Moreover
%e9 ###
%
\begin{equation}\label{condkappa}
\frac{1}{1+(1-\lambda_l) (\kappa^{-1}(t_l^{-1}-t_{l-1}^{-1})-1
)}<\theta_l\leq1,\qquad
l=1,\ldots,K.
\end{equation}
\end{assumption}

\eject
\begin{remark}
\begin{longlist}
\item[(1)]The drift and minorization conditions (\ref{mino})--(\ref
{drift}) of Assumption~\ref{assum1} can be checked for many practical
examples. If each $P^{(l)}$ is a Random Walk Metropolis kernel or a
Metropolis Adjusted Langevin kernel then (\ref{mino}) and (\ref
{drift}) are known to hold under some regularity conditions on the
energy function $E$ (see \cite{jarnerethansen98,atchade05}). In these
cases, it is always possible to choose $\kappa$ small enough to
satisfy $0<\kappa< (\frac{1}{t_l}-\frac{1}{t_{l-1}} )$.
\item[(2)] The condition (\ref{condkappa}) is a technical condition
that quantifies the idea that the rate of resampling $1-\theta_l$
should not be too large. It is needed to guarantee that the geometric
drift condition (\ref{drift}) on $P^{(l)}$ transfers to kernels of the
type $P_\mu^{(l)}$ that drive the EE sampler.
%Note that if $P^{(l)}$ satisfies a geometric drift condition (
%geometric drift condition with respect to $V^\alpha(x)=c^\alpha e^{
%words, condition (\ref{condkappa}) is not restrictive.
\end{longlist}
\end{remark}

%s3.2 ###
\subsection{Law of large numbers}
We consider an arbitrary pair $\{(X_n^{(l-1)},X_n^{(l)}),\break n\geq0\}$.
We will show that under Assumption~\ref{assum1}, if $\{X_n^{(l-1)},
n\geq0\}$ satisfies a strong law of large numbers, then so does $\{
X_n^{(l)}, n\geq0\}$. Then we use the fact that $\{X_n^{(0)}, n\geq0\}
$ is an ergodic Markov chain to derive a law of large numbers for any
$\{X_n^{(l)}, n\geq0\}$.

\begin{theorem}\label{thm1}Assume Assumption~\ref{assum1} holds and let
$\beta\in[0,1)$. Let $f\dvtx (\M,\B_\M)\times(\X,\B)\to\Real
$ be a measurable function such that
%e10 ###
%
\begin{equation}\label{eq1keyergothmEE}\sup_{\nu\in\M}\abs{f_\nu
}_{V^\beta}<\infty.
\end{equation}
Suppose that there exists a finite constant $C$ such that for any $\nu
,\mu\in\M$,
%e11 ###
%
\begin{equation}\label{eq2keyergothmEE}\abs{f_\nu-f_\mu}_{V^\beta
}\leq C\Vert\nu-\mu\Vert_{V^\beta}.
\end{equation}
Suppose also that for any $h\in L^\infty_{V^\beta}$,
%e12 ###
%
\begin{equation}\frac{1}{n}\sum_{k=1}^nh\bigl(X_k^{(l-1)}\bigr)\To
\pi^{(l-1)}(h),\qquad \p\mbox{-a.s.  as } n\to\infty,
\end{equation}
and that there exists $\D\in\F$, $\p(\D)=1$ such that for each
sample path $\omega\in\D$, $f_{\mu^{(l-1)}_{n}}(x)(\omega)$
converges to $f_{\pi^{(l-1)}}(x)$ as $n\to\infty$ for all $x\in\X$.
Then
%e13 ###
%
\begin{equation}\frac{1}{n}\sum_{k=1}^nf_{\mu
^{(l-1)}_{k-1}}\bigl(X_k^{(l)}\bigr)\To\pi^{(l)} \bigl(f_{\pi^{(l-1)}} \bigr),\qquad \p
\mbox{-a.s. as } n\to\infty.
\end{equation}
\end{theorem}
\begin{pf}See Section~\ref{proofthm1}.
\end{pf}

The following corollary is then immediate.
\begin{corollary}
Assume Assumption~\ref{assum1} holds and suppose that $\{X_n^{(0)},\break n\geq0\}
$ is a $\phi$-irreducible aperiodic Markov chain with invariant
distribution $\pi^{(0)}$ and $\pi^{(0)}(V)<\infty$. Let $f\in
L^\infty_{V^{\beta}}$, $\beta\in[0,1)$. Then for any $l\in\{
1,\ldots,K\}$,
%e14 ###
%
\begin{equation}
\frac{1}{n}\sum_{i=1}^{n}f\bigl(X_i^{(l)}\bigr)\To\pi
^{(l)}(f),\qquad \p\mbox{-a.s.  as } n\to\infty.
\end{equation}
\end{corollary}

%s3.3 ###
\subsection{Central limit with a random centering}\label{cltdet}

We now turn to central limit theorems. It can be shown that the kernel
$P_\mu^{(l)}$ admits a unique invariant distribution $\pi_\mu
^{(l)}$. Since the conditional distribution of $X_n^{(l)}$ given $\F
_{n-1}$ is $P_{\mu_{n-1}^{(l-1)}}^{(l)}$, it is natural to consider a
central limit theorem for $\sum_{k=1}^nf(X_k^{(l)})$ in which
$f(X_k^{(l)})$ is centered around $\pi_{\mu_{n-1}^{(l-1)}}^{(l)}(f)$.
This is done in the next theorem. $\Rightarrow$ denotes weak
convergence and $\mathcaligr{N}(\mu,\sigma^2)$ denotes the Gaussian
distribution on $\Real$ with mean $\mu$ and variance~$\sigma^2$.
\begin{theorem}\label{thm2}
Assume Assumption~\ref{assum1} holds.
Let $f\in L^\infty_{V^{\beta}}$, $\beta\in[0,1/2)$ be such that
$\pi^{(l)}(f)=0$. Define
%e15 ###
%
\begin{equation}\label{asympvarEE}
\sigma_l^2(f):=\pi^{(l)}(f^2)+2\sum_{k=1}^\infty\int_\X\pi
^{(l)}(dx)f(x) \bigl[K^{(l)} \bigr]^kf(x),
\end{equation}
where $K^{(l)}$ is given by (\ref{Plimee}). Assume that $\sigma
_l^2(f)>0$. Then there exists a random sequence $\{\pi_n^{(l)}(f)\}$,
$\pi_n^{(l)}(f)\to\pi^{(l)}(f)$ (almost surely) as $n\to\infty$
such that
%e16 ###
%
\begin{equation}
\frac{1}{\sqrt{n}\sigma_l(f)}\sum_{k=1}^{n}
\bigl[f(X_k^{(l)})-\pi_k^{(l)}(f) \bigr]\Rightarrow\mathcaligr{N}(0,1)\qquad \mbox{as } n\to\infty.
\end{equation}
\end{theorem}
\begin{pf}See Section \ref{proofthm2}.
\end{pf}

%s3.4 ###
\subsection{Central limit theorem with a deterministic
centering}\label{secthm3}
We now derive a central limit theorem for $\sum_{k=1}^nf(X_k^{(l)})$
around $\pi^{(l)}(f)$ which gives more insight in the efficiency of
the method as a Monte Carlo sampler from $\pi^{(l)}$. We restrict
ourselves to the case where $l=1$; that is, we only consider the pair
$\{(X_n^{(0)},X_n^{(1)}), n\geq0\}$. Moreover, we assume in this
section that $\X$ is a compact subset of $\Real^d$ (equipped with its
Euclidean metric). More precisely:

\renewcommand{\theassumption}{(A1$'$)}
\begin{assumption}\label{assum'} $\X$ is a compact subset of
$\Real^d$. For $l=0,1$, there exist an integer $n_0>0$, a constant
$\eps_l \in(0,1]$ a probability measure $\phi_l$ such that for $x
\in\X$ and $A\in\B$,
%e17 ###
%
\begin{equation}\label{minounif}
\bigl[P^{(l)} \bigr]^{n_0}(x,A)\geq\eps_l\phi_l(A) .
\end{equation}
\end{assumption}

Let $\C(\X,\Real)$ be the space of all continuous functions from $\X
\to\Real$. We endowed $\C(\X,\Real)$ with the uniform metric
$|f|_\infty:=\sup_{x\in\X}|f(x)|$ and its Borel $\sigma$-algebra.
Let $\operatorname{Lip}(\X,\Real)$ be the subset of Lipschitz functions of
$\C(\X,\Real)$ [we say that $f\dvtx\X\to\Real$ is Lipschitz if there
exists $C<\infty$ such that for any $x,y\in\X$, $|f(x)-f(y)|\leq C|x-y|$].

For $f\dvtx \X\to\Real$ bounded measurable, define the function
\[
U(x)=U_f(x):=\sum_{j\geq0} \bigl(P^{(1)}_{\pi^{(0)}} \bigr)^jf(x),
\]
the solution to the Poisson equation for $f$ and $P^{(1)}_{\pi
^{(0)}}$. To simplify\vspace*{1pt} the notations, we omit the dependence of $U$ on
$f$. Notice that $P^{(1)}_{\pi^{(0)}}$ is the limiting kernel in the
EE sampler, denoted $K^{(1)}$ in (\ref{Plimee}). Clearly, Assumption~\ref{assum'}
implies as shown in Lemma \ref{DrMinoEE} below that the kernel $P_\mu
^{(1)}$ is also uniformly ergodic, uniformly in $\mu$. In particular
$|U|_\infty<\infty$. We assume that the function $U$ is Lipschitz
whenever $f$ is Lipschitz:
%e18 ###
%
\begin{equation}\label{LipAssump}
f\in\operatorname{Lip}(\X,\Real)
\qquad\mbox{implies that}\quad \sum_{j\geq0} \bigl(P^{(1)}_{\pi^{(0)}} \bigr)^jf \in
\operatorname{Lip}(\X,\Real).
\end{equation}

We comment on (\ref{LipAssump}) below. Let $f\in\C(\X,\Real)$ such
that $\pi^{(1)}(f)=0$. Consider the partial sum $S_n=\sum_{k=1}^n
f(X_k^{(1)})$. Since $U$ satisfies the Poisson equation $U-P_{\pi
^{(0)}}^{(1)} U=f$, we can rewrite $S_n$ as
\begin{eqnarray*}S_n
&=&
\sum_{k=1}^nU\bigl(X_k^{(1)}\bigr)-P_{\pi
^{(0)}}^{(1)}U\bigl(X_k^{(1)}\bigr)
\\
&=& M_n+\sum_{k=1}^nP_{\mu_k^{(0)}}^{(1)} U\bigl(X_k^{(1)}\bigr)- P_{\pi
^{(0)}}^{(1)} U\bigl(X_k^{(1)}\bigr) + \eps_n^{(1)},
\end{eqnarray*}
where $M_n=\sum_{k=1}^nU(X_k^{(1)})-P_{\mu
_{k-1}^{(0)}}^{(1)}U(X_{k-1}^{(1)})$ is a martingale and\vspace*{-3pt} $\eps
_n^{(1)}=P_{\mu_{0}^{(0)}}^{(1)}\times\break U(X_{0}^{(1)})-P_{\mu
_{n}^{(0)}}^{(1)}U(X_{n}^{(1)})$.

We introduce the function
%e19 ###
%
\begin{eqnarray}\label{Hxfun}H_x(y)
&:=&
T^{(1)}(y,x,U)-R^{(1)}(x,U)\nonumber\\[-8pt]\\[-8pt]
&&\hspace*{-10.8pt}=\int
T^{(1)}(y,x,dz)U(z)-\int\pi^{(0)}(dy)\int T^{(1)}(y,x,dz)U(z).\nonumber
\end{eqnarray}
Since $P_\mu^{(1)}(x,dz)=\theta_1P^{(1)}(x,dz)+(1-\theta_1)\int\mu
(dy)\int T^{(1)}(y,x,dz)$, we have
\[
P_\mu^{(1)} U(x)-P_{\pi^{(0)}}^{(1)} U(x)=(1-\theta_1)\int\mu(dy) H_x(y),
\]
so that we can rewrite $S_n$ as
\begin{eqnarray*}
S_n &=&
M_n+(1-\theta_1)\sum_{k=1}^n\frac{1}{k}\sum
_{j=1}^kH_{X^{(1)}_k}\bigl(X_j^{(0)}\bigr) + \eps_n^{(1)}\\
&=& M_n+(1-\theta_1)\sum
_{k=1}^n\frac{1}{\sqrt{k}}\eta_k\bigl(X^{(1)}_k\bigr) + \eps_n^{(1)},
\end{eqnarray*}
where $\eta_n$ is the random field
\[
\eta_n(x):=n^{-1/2}\sum_{k=1}^nH_x\bigl(X_k^{(0)}\bigr).
\]
We will see that $\eta_n$ is a $\C(\X,\Real)$-valued random
element. To describe its asymptotic behavior we introduce the function
\[
U_x^{(0)}(y)=\sum_{j\geq0} \bigl[P^{(0)} \bigr]^jH_x(y),
\]
where for a kernel $Q$, $QH_x(y)=\int Q(y,dz)H_x(z)$ and the covariance function
%e20 ###
%
\begin{equation}\label{Gamma}
\Gamma(x,y)=\int\bigl[U_x^{(0)}(z)U_y^{(0)}(z)- \bigl(P^{(0)}U_x^{(0)}(z) \bigr)
\bigl(P^{(0)} U_y^{(0)}(z) \bigr) \bigr]\pi^{(0)}(dz).
\end{equation}
If $f,g\in\C(\X,\Real)$, with an abuse of notation we will also
write $\Gamma(f,g)$ for the quantity
\[
\Gamma(f,g)=\int\bigl[U_f^{(0)}(z)U_g^{(0)}(z)- \bigl(P^{(0)}U_f^{(0)}(z) \bigr)
\bigl(P^{(0)} U_g^{(0)}(z) \bigr) \bigr]\pi^{(0)}(dz),
\]
where $U_f^{(0)}(x)=\sum_{j\geq0} [P^{(0)} ]^jf(x)$.

%From standard results on the central limit theorem for uniformly
%ergodic Markov chains, the finite dimensional distribution of $\eta_n$
%converges weakly to the finite dimensional distribution of a mean zero
%Gaussian process $G$ on $\X$ with sample paths in $\C(\X,\Real)$ and
%covariance function $\Gamma$. We will actually show that $\eta_n$
%converges weakly to $G$ in $\C(\X,\Real)$ and we prove the following
%result.

\begin{theorem}\label{thm3}
Assume Assumption~\ref{assum'} and (\ref{LipAssump}) hold and suppose that $E\in\operatorname{Lip}(\X
,\Real)$. Let $f\in\operatorname{Lip}(\X,\Real)$ such that $\pi
^{(1)}(f)=0$. Then
%e21 ###
%
\begin{equation}\frac{1}{\sqrt{n}}\sum
_{k=1}^{n}f\bigl(X_k^{(1)}\bigr)\Rightarrow\mathcaligr{N} \bigl(0,\sigma_\star
^2(f)+4(1-\theta_1)^2\Gamma(\bar g,\bar g) \bigr)\qquad \mbox{as } n\to\infty,
\end{equation}
where $\bar g(\cdot):=\int\pi^{(1)}(dx)H_x(\cdot)$ and
%e22 ###
%
\begin{equation}
\sigma_\star^2(f):=\pi^{(1)}(f^2)+2\sum_{k=1}^\infty\int_\X\pi
^{(1)}(dx)f(x) \bigl(P_{\pi^{(0)}}^{(1)} \bigr)^kf(x).
\end{equation}
\end{theorem}
\begin{pf}See Section \ref{proofthm3}.
\end{pf}

Notice from (\ref{Hxfun}) that $\bar g(\cdot)=\int\pi
^{(1)}(dx)T^{(1)}(\cdot,x,U)-\int\pi^{(0)}(dz)\int\pi
^{(1)}(dx)\times\break T^{(1)}(z,x,U)$. Thus Theorem \ref{thm3} shows that the
asymptotic variance of the EE sampler is the sum of the asymptotic
variance in estimating $\pi^{(1)}(f)$ as if the limiting kernel
$P^{(1)}_{\pi^(0)}$ is known [the term $\sigma_\star^2(f)$] plus the\vspace*{-1pt}
asymptotic in using the chain $\{X_n^{(0)}, n\geq0\}$ to estimate the
expectation under $\pi^{(0)}$ of the function $\int\pi
^{(1)}(dx)T^{(1)}(\cdot,x,U)$. In their analysis \cite
{doucetetdelmoral08b} arrive at a similar CLT for interacting MCMC
algorithms. Notice also that $U(x)=\sum_{j\geq0} (P_{\pi
^{(0)}}^{(1)})^jf(x)$. Thus in most cases, the function $\int\pi
^{(1)}(dx)T^{(1)}(\cdot,x,U)$ will typically take large values and the
asymptotic variance in estimating its expectation will also tend to be
large particularly if the kernel $P^{(0)}$ mixes poorly. Theorem \ref
{thm3} thus suggests that for the EE sampler to be effective in
practice it is important that the initial chain $\{X_n^{(0)}, n\geq0\}
$ enjoys a very fast mixing.

A remaining question is to know whether $n^{-1}\E[ (\sum_{k=1}^n
f(X_k^{(1)}) )^2 ]$ converges to $\sigma_\star^2(f) + 4(1-\theta
_1)^2\Gamma(\bar g,\bar g)$. Unfortunately the answer is no in general
as shown by the following example:

\begin{proposition}\label{propAsympVar}
Assume Assumption~\ref{assum'} holds. Suppose that $P^{(0)}=P^{(1)}=P$ and $\pi^{(0)}=\pi
^{(1)}=\pi$. Let $f\dvtx \X\to\Real$ be a bounded measurable function
such that $\pi(f)=0$. Then
\[
\lim_{n\to\infty} n^{-1}\E\Biggl[ \Biggl(\sum_{k=1}^n f\bigl(X_k^{(1)}\bigr) \Biggr)^2
\Biggr]=\sigma_\star^2(f) + 2(1-\theta_1)^2\Gamma(\bar g,\bar g).
\]
In the present case $\bar g(x)=U(x)=\sum_{j\geq0}\theta_1^j P^j
f(x)$ and
\[
\sigma^2_\star(f)=\pi(|f|^2)+2\sum_{k=1}^\infty\theta_1^k\int\pi
(dx)f(x)P^kf(x).
\]
%(P^jf,P^jf )+2\sum_{0\leq%i<j}\theta_1^{i+j}\Gamma(P^if,P^jf ).\]
\end{proposition}
\begin{pf}See Section \ref{proofpropAsympVar}.
\end{pf}

\begin{remark}
Assumption (\ref{LipAssump}) can often be easily checked. Indeed, we
have $U(x)=f(x)+P_{\pi^{(0)}}^{(1)} U(x)$, where $P_{\pi
^{(0)}}^{(1)}=\theta_1 P^{(1)} +(1-\theta_1)R^{(1)}$, where $R^{(1)}$
is the independent Metropolis--Hastings algorithm with target $\pi
^{(1)}$ and proposal $\pi^{(0)}$. Let us assume that $P^{(1)}$ is also
a Metropolis--Hastings kernel with target $\pi^{(1)}$ and proposal
$q(x,y)$. Denote $\alpha(x,y)$ [resp. $\bar\alpha(x,y)$] the
acceptance probability of $P^{(1)}$ [resp. $R^{(1)}$], and denote
$a(x):=\int\alpha(x,y)q(x,y)\,dy$ [resp. $a(x):=\int\alpha(x,y)\pi
^{(0)}(y)\,dy$] the average acceptance probability at $x$ for $P^{(1)}$
[resp. for $R^{(1)}$]. Then we have
\begin{eqnarray*}
&&\hspace*{-5pt}U(x) \bigl(1-\theta_1\bigl(1-a(x)\bigr)-(1-\theta_1)\bigl(1-\bar a(x)\bigr) \bigr)\\
&&\hspace*{-5pt}\qquad=f(x)+\theta_1\int\alpha(x,y)q(x,y)U(y)\,dy +(1-\theta_1)\int\bar
\alpha(x,y)\pi^{(0)}(y)U(y)\,dy.
\end{eqnarray*}
Thus if $\pi^{(0)},\pi^{(1)}$ and $q$ such that $a$ and $\bar a$
remains bounded away from\break $0$ and the integral operators $h\to\int
\alpha(x,y)q(x,y)h(y)\,dy$ and $h\to\break\int\bar\alpha(x,y)\pi
^{(0)}(y)h(y)\,dy$ transform bounded measurable functions into Lipschitz
functions, then (\ref{LipAssump}) hold. For example, if $\pi
^{(0)},\pi^{(1)}$ and $q$
are all positive on $\X$ and of class $\C^1$ then (\ref{LipAssump}) hold.
\end{remark}

%
%To give some more insight, suppose that and take In this case the EE
%algorithm consists in running two parallel chains with the same
%%target distribution $\pi^{(1)}$ with the second chain resampling from
%the first. Then the limiting kernel of the second chain is %$\theta_1
%P+ (1-\theta_1)\pi$ and the term $\sigma^2_\star(f)$ becomes
%which can be drastically smaller than the $\Gamma(f,f)$. But in this
%case we have $\int\pi^{(1)}(dx)T^{(1)}(z,x,U)=U(z)=\sum_{k\geq%0}
%chain $\{X_n^{(0)}, n\geq0\}$ is
%which can be substantial larger than $\Gamma(f,f)$ particularly if the
%mixing of the kernel $P$ is poor.
%
%Even in the case $\pi^{(0)}\neq\pi^{(1)}$, the function $\int
%and can be "typically large" and as such the term $\Gamma(\bar g,\bar
%g)$ can be substantial.
%

\begin{remark}
The result developed above relies heavily on the Lipschitz continuity
assumption. Under that assumption, we show that the stochastic process
$\{\eta_n, n\geq0\}$ lives in the Polish space $\C(\X,\Real)$
which allows us to use the standard machinery of weak convergence in
Polish spaces. If $f$ is only assumed measurable the theorem above no
longer hold. But a similar result can still be obtained using weak
convergence techniques in nonseparable metric spaces. But we do not
pursue this here.
\end{remark}

%We discuss the efficiency of IR-MCMC as a consequence of Thereom
%of Theorem \ref{thm2}. The $l$-th limiting kernel of IR-MCMC is
%$K^{(l)}=\theta_lP^{(l)}+(1-\theta_l)\pi^{(l)}$. Let $f$ as in Theorem
%interesting to note that $\sigma_l^2(f)$ in (\ref{asympvarIR}) is
%precisely the asymptotic variance for $f$ in the central limit theorem
%for the Markov chain with transition kernel $K^{(l)}$. More precisely,
%if $\{Y_n^{(l)}\}$ is a Markov chain with transition kernel $K^{(l)}$,
%then $\{Y_n^{(l)}\}$ is geometrically ergodic and $
%where $\sigma_l^2(f)$ is given by (\ref{asympvarIR}).
%Denote $\mathcaligr{N}(\mu,\sigma^2)$ the normal distribution with
%mean $\mu$ and variance $\sigma^2$.
%
%For $n$ large it follows from Theorem \ref{thm1} that
%where
%Therefore the precision of the Monte Carlo estimate $
%
%As we will see below, the term $\pi_n^{(l)}(f)$ can be written
% as an empirical sum based on $\{X_k^{(l-1)}, k\leq n-1\}$ for some
%function $R_{\theta,f}$. Thus this term is typically of the order $1/
%)^2/n^2+2\mathbb{C}\mbox{ov} (\sum_{k=1}^n\pi_k^{(l)}(f)/n,
%Mean Square Error of $\sum_{k=1}^nf(X_k^{(l)})/n$ is not $
%than $\sigma_l^2(f)$ as the simulation example shows.
%

%s3.5 ###
\subsection{An illustrative example}\label{ex1}

Consider the following example. Suppose that we want to sample from the
bivariate normal distribution $\mathcaligr{N} (0,\Sigma)$, with
covariance matrix
\[
\Sigma= \left[
\matrix{0.96& 2.44\cr
2.44& 7.04
}
\right].
\]
For this problem, we compare a Random Walk Metropolis (RWM) algorithm,
the EE sampler, the MCMC algorithm based on the limiting kernel of EE
sampler (call it limit EE sampler), IR-MCMC and the MCMC algorithm
based on the limiting kernel of IR-MCMC (limit IR-MCMC sampler).

For the RWM sampler, the proposal kernel is $\mathcaligr{N} (0,I_2 )$,
where $I_2$ is the $2$-dimensional identity matrix. For the adaptive
chains, we use four chains with $\pi^{(0)}=\pi^{1/10}$, $\pi
^{(1)}=\pi^{1/5}$, $\pi^{(2)}=\pi^{1/2}$ and $\pi^{(3)}=\pi$. We
take $\theta_l=\theta=0.5$ and $P^{(l)}$ is taken to be a RWM
algorithm with target $\pi^{(l)}$ and proposal $\mathcaligr{N} (0,I_2
)$. It can be checked that Assumption~\ref{assum1} holds for this
problem. We simulate each of the five samplers for $N=10\mbox{,}000$
iterations. We compare the samplers on their mean square errors (MSE)
in estimating the first two moments of the two components of the
distribution $\pi$. We calculate the MSEs by repeating the simulations
$100$ times. The results are reported in Table~\ref{table1}.

From these results we see (as expected) that the limit EE sampler is
$3$ to $25$ times more efficient than the RWM sampler, and the limit
IR-MCMC sampler is $15$ to $50$ more efficient than the RWM sampler.
But IR-MCMC itself is hardly more efficient than the RWM sampler. If we
take the computation times into account, it becomes hard to make the
case that any of these adaptive sampler is better than the plain RWM.
Similar conclusions can be drawn for the EE sampler.
%
%It is important to note that this poor performance of IR-MCMC is not
%due to a poor choice of the interpolating distributions $\pi^{(t)}$.
%With the choice given above, the estimated efficiencies $

%t1 ###
%
\begin{table}[b]
\caption{Mean square error and ratios (with respect to the RWM
sampler) for IR-MCMC, limit IR-MCMC, EE and limit EE. Based on \textit{100}
replications of \textit{10,000} iterations of each sampler}\label{table1}
\begin{tabular*}{\textwidth}{@{\extracolsep{\fill}}lcd{2.4}d{2.4}d{2.4}d{2.4}@{}}
\hline
& & \multicolumn{1}{c}{$\mathbf{\E(X_1)}$} & \multicolumn{1}{c}{$\mathbf{\E(X_2)}$}
& \multicolumn{1}{c}{$\mathbf{\E(X_1^2)}$}
& \multicolumn{1}{c@{}}{$\mathbf{\E(X_2^2)}$}\\
\hline
RWM &MSE& 0.0099&0.0803 &0.0091 &0.5525 \\
& Ratios & 1.0 & 1.0& 1.0& 1.0\\[3pt]
{IR-MCMC} & MSE & 0.0098 & 0.0774 & 0.0047 & 0.2962\\
& Ratios & 1.00 & 1.04 & 1.95 & 1.87\\[3pt]
{Limit IR-MCMC} & MSE & 0.0002& 0.0017 & 0.0006 & 0.0296 \\
& Ratios & 48.43 & 46.20 & 14.18 & 18.66\\[3pt]
{EE} & MSE & 0.0057 & 0.0435 & 0.0045 & 0.2810\\
& Ratios & 1.74 & 1.84 & 2.02 & 1.97\\[3pt]
{Limit EE} & MSE & 0.0004& 0.0030 & 0.0034 & 0.1966 \\
& Ratios & 25.99 & 26.36 & 2.67 & 2.81\\
\hline
\end{tabular*}
\end{table}

\eject
%s4 ###
\section{Proofs}\label{proof}
%Throughout the proof we will use the letter $C$ to denote a generic
%finite constant whose actual value might varies from one appearance to
%another. The EE sampler generates the nonhomogeneous Markov chain on $
%given by
%(dx^{(0)}',\ldots,dx^{(K)}',d\mu^{(0)}'\ldots,d\mu^{(K-1)}' ) )\\
%=P^{(0)}(x^{(0)},dx^{(0)}')\prod_{l=1}^KP^{(l)}_{\mu^{(l-1)}}
%(x^{(l)},dx^{(l)}' )\prod_{l=0}^{K-1}\delta_{\{\mu^{(l)}+n^{-1}(
%where
%and
% We denote $\{\F_n\}$, $\PP$ and $\PE$ respectively the natural
%filtration, probability distribution and expectation operator of the
%process.
%
%The main ingredient of the proof is martingale approximation (
%%martingale
%approximation if there exist random processes $\{M_n\}$ and $\{R_n\}$
%such that: (i) $Y_n=M_n+R_n$ for all $n\geq1$ (ii) $\{M_n,\F_n\}$ is
%a square-integrable martingale and $\{R_n\}$ is negligible in some
%sense (for example $\sup_n\E[\abs{R_n}^\alpha]<\infty$ for some $
%show that such martingale approximation exists for each of the $
%
%The strategy to the proof of Theorem \ref{thm2} is similar to that of
%Theorem \ref{thm1}. We will show the $l$-th chain inherits the
%ergodicity of the the $(l-1)$-th chain. And since the initial chain is
%an ergodic Markov chain, we will conclude that the entire process is
%ergodic. In particular, a martingale approximation will be made
%available to prove the central limit theorem. But the details of the
%proofs here are less straightforward and rely heavily on the Polish
%assumption. The key result is Theorem \ref{keyergothmEE}. We start
%with some preliminary lemmas on kernels of the form $P^{(l)}_\nu$.

%s4.1 ###
\subsection{Preliminary results on kernels of the form $P^{(l)}_\nu$}

For a probability measure~$\nu$ and $l=1,\ldots,K$, let $P_\nu
^{(l)}$ as in (\ref{Pmu}) with $\omega^{(l)}\equiv1$ and $T^{(l)}$
as in (\ref{Tlee}). The following lemma shows that $P^{(l)}_{\nu}$
satisfies a drift and a minorization conditions with constant that
actually do not depend on $\nu$.

\begin{lemma}\label{DrMinoEE}
Assume Assumption~\ref{assum1} holds. Then
there exists $\lambda_l'\in(0,1)$ that does not depend on $\nu$ such
that for $x \in\X$ and $A\in\B$:
%e23 ###
%
\begin{equation}\label{mino2}
\bigl[P^{(l)}_{\nu} \bigr]^{n_0}(x,A)\geq\theta
_l\eps_l\phi_l(A)\mathbf{1}_{C_l}(x)
\end{equation}
and
%e24 ###
%
\begin{equation}\label{drift2}
P^{(l)}_{\nu}V(x)\leq\lambda'_l V(x)+b_l\mathbf{1}_{C_l}(x) ,
\end{equation}
where $C_l$, $\phi_l$, $b_l$, $\eps_l$ and $V$ are as in
Assumption~\ref{assum1}.
\end{lemma}
\begin{pf}
We have $P_\nu^{(l)}\geq\theta_lP^{(l)}$. Therefore (\ref{mino2})
follows from the minorization condition (\ref{mino}).

Define $\delta_l= (\kappa^{-1} (t_l^{-1}-t_{l-1}^{-1} )-1 )^{-1}$. We
will show that
%e25 ###
%
\begin{equation}\label{lemee1eq1}\int\nu(dy)T^{(l)}(y,x,V)\leq
(1+\delta_l)V(x).
\end{equation}
Given the drift condition (\ref{drift}), this will imply
\begin{eqnarray*}
P^{(l)}_{\nu}V(x)
&\leq&
\bigl(\theta_l\lambda+(1-\theta
_l)(1+\delta_l) \bigr) V(x)+b_l\mathbf{1}_{C_l}(x)\\
&\leq&
\lambda_l'V(x)+b_l\mathbf{1}_{C_l}(x),
\end{eqnarray*}
where $\lambda_l'=\theta_l\lambda+(1-\theta_l)(1+\delta_l)\in
(0,1)$ by the condition on $\kappa$ in Assumption~\ref{assum1}.

Observe that $r^{(l)}(x)=e^{-E(x)(t_l^{-1}-t_{l-1}^{-1})}$,
$t_l^{-1}-t_{l-1}^{-1}>0$ and $V(x)=ce^{\kappa E(x)}\geq1$, $\kappa
\in(0,1)$. This implies that $r^{(l)}(y)/r^{(l)}(x)\geq1$ if and only
if $E(y)\leq E(x)$. Denote $\mathcaligr{A}(x)=\{y\in\X\dvtx E(y)\leq
E(x)\}$ and $\mathcaligr{R}(x)=\{y\in\X\dvtx E(y)> E(x)\}$. Then we have
\begin{eqnarray*}
&&\int\nu(dy)T^{(l)}(y,x,V)\\
&&\qquad=
\int_{\mathcaligr{A}(x)}\nu
(dy)T^{(l)}(y,x,V)+\int_{\mathcaligr{R}(x)}\nu(dy)T^{(l)}(y,x,V)\\
&&\qquad=
\int_{\mathcaligr{A}(x)}\nu(dy)V(y)+\int_{\mathcaligr{R}(x)}\nu
(dy)\frac{r^{(l)}(y)}{r^{(l)}(x)}V(y) \\
&&\qquad\quad{}+ V(x)\int_{\mathcaligr
{R}(x)}\nu(dy) \biggl(1-\frac{r^{(l)}(y)}{r^{(l)}(x)} \biggr),\\
&&\qquad=
\int_{\mathcaligr{A}(x)}\nu(dy)V(y)+V(x)\int_{\mathcaligr
{R}(x)}\nu(dy)\\
&&\qquad\quad{}+\int_{\mathcaligr{R}(x)}\nu(dy)\frac
{r^{(l)}(y)}{r^{(l)}(x)} \bigl(V(y)-V(x) \bigr)\\
&&\qquad\leq
V(x)+V(x)\int_{\mathcaligr{R}(x)}\nu(dy)\frac
{r^{(l)}(y)}{r^{(l)}(x)} \biggl(\frac{V(y)}{V(x)}-1 \biggr)\\
&&\qquad=
V(x) \biggl[1+\int_{\mathcaligr{R}(x)}e^{- (E(y)-E(x) ) (1/t_l-1/t_{l-1}
)}\\
&&\qquad\quad\hspace*{70pt}{}\times \bigl(e^{\kappa(E(y)-E(x) )}-1 \bigr)\nu(dy) \biggr]\\
&&\qquad\leq
V(x)\frac{\kappa}{1/t_{l}-1/t_{l-1}-\kappa}.
\end{eqnarray*}
In the last line we use the following inequality: for $0<x<y$:
$e^{-y}(e^x-1)\leq x/(y-x)$.
\end{pf}

From Lemma \ref{DrMinoEE}, we deduce that for any probability measure
$\nu$, $P^{(l)}_{\nu}$ has an invariant distribution $\pi^{(l)}_{\nu
}$ such that
%e26 ###
%
\begin{equation}\pi^{(l)}_{\nu}(V)\leq b_l.
\end{equation}
See \cite{meynettweedie93}, Theorems 15.0.1 and 14.3.7. The lemma
also implies that for any $\beta\in(0,1]$, there exist constants
$C_\beta<\infty$ and $\rho_\beta\in(0,1)$ that does not depend on
$\nu$ such that
%e27 ###
%
\begin{equation}\label{ratePn}
\big\Vert \bigl[P^{(l)}_{\nu} \bigr]^k(x,\cdot
)-\pi^{(l)}_{\nu}(\cdot)\big\Vert_{V^\beta}\leq C_\beta\rho_\beta
^kV^\beta(x),\qquad k\geq0, x\in\X.
\end{equation}
See, for example, \cite{baxendale05} for a proof.
The following lemma holds.
\begin{lemma} \label{stabEE}Fix $\beta\in[0,1]$ and $\mu$ and $\nu
$ two probability measures on $(\X,\B)$
%e28 ###
%
\begin{equation}\nnnorm{P^{(l)}_{\mu}-P^{(l)}_{\nu}}_{V^\beta}\leq
2\Vert\mu-\nu\Vert_{V^\beta}.
\end{equation}
\end{lemma}
\begin{pf}
For $f\in L^\infty_{V^\beta}$ such that $\abs{f}_{V^\beta}\leq1$,
we have
\[
P^{(l)}_{\mu}f(x)-P^{(l)}_{\nu}f(x)=(1-\theta_l)\int T^{(l)}(y,x,f)
\bigl(\mu(dy)-\nu(dy) \bigr),
\]
where $T^{(l)}(y,x,f)= \min(1,\frac{r_l(y)}{r_l(x)} ) (f(y)-f(x)
)+f(x)$. Therefore
\begin{eqnarray*}
&&\frac{P^{(l)}_{\mu}f(x)-P^{(l)}_{\nu}f(x)}{(1-\theta_l)V^\beta
(x)}\\
&&\qquad=\int\frac{\min(1,r^{(l)}(y)/r^{(l)}(x) ) (f(y)-f(x)
)}{V^\beta(x)V^\beta(y)}V^\beta(y) \bigl(\mu(dy)-\nu(dy) \bigr).
\end{eqnarray*}

Now for $\abs{f}_{V^\beta}\leq1$, $\abs{\frac{\min(1,r^{(l)}(\cdot)/r^{(l)}(x) ) (f(\cdot)-f(x) )}{V^\beta(x)V^\beta
(\cdot)}V^\beta(\cdot)}_{V^\beta}\leq2$ for all $x\in\X$. Therefore
\begin{eqnarray*}
&&\bigg\vert\int\frac{\min(1,r^{(l)}(y)/r^{(l)}(x) ) (f(y)-f(x) )}
{V^\beta(x)V^\beta(y)}V^\beta
(y) \bigl(\mu(dy)-\nu(dy) \bigr)\bigg\vert\\
&&\qquad \leq
2\sup_{\abs{f}_{V^\beta}\leq1}\bigg\vert\int f(y) \bigl(\mu(dy)-\nu(dy) \bigr)\bigg\vert\\
&&\qquad = 2\Vert\mu-\nu\Vert_{V^\beta}.
\end{eqnarray*}
\upqed
\end{pf}

For $l\in\{1,\ldots,K\}$, define the kernel
\[
N^{(l)}_\mu f(x)=\int\mu(dy)f(y)\min\biggl(1,\frac
{r^{(l)}(y)}{r^{(l)}(x)} \biggr),\qquad x\in\X.
\]

\begin{lemma}\label{lemunifcont} Let $\mu$  be a probability
measure on $(\X,\B)$. For $x_1,x_2\in\X$, and $f\in L^\infty
_{V^\beta}$, $\beta\in[0,1]$
%e29 ###
%
\begin{eqnarray}\label{unifcont}
&&\big\vert N^{(l)}_\mu f(x_1)-N^{(l)}_\mu
f(x_2)\big\vert\nonumber\\[-8pt]\\[-8pt]
&&\qquad\leq\abs{f}_{V^\beta}\big\vert e^{\tau E(x_1)}-e^{\tau
E(x_2)}\big\vert \bigg\vert\int\mu(dy)e^{-(\tau-\kappa\beta)E(y)}\bigg\vert\nonumber
\end{eqnarray}
with $\tau=1/t_l-1/t_{l-1}$ and $\kappa$ as in Assumption~\ref{assum1}.
\end{lemma}
\begin{pf}
Fix $x_1$ and $x_2$ and define\vspace*{-2pt} $\Delta(y)=V^\beta(y)\abs{\min
(1,\frac{r^{(l)}(y)}{r^{(l)}(x_1)} )-\min(1,\break\frac
{r^{(l)}(y)}{r^{(l)}(x_2)} )}$. On $r^{(l)}(y)\geq\max
(r^{(l)}(x_1),r^{(l)}(x_2))$, $\Delta(y)=0$. On $r^{(l)}(x_1)\leq
r^{(l)}(y)\leq r^{(l)}(x_2)$,
\begin{eqnarray*}
\Delta(y)
&=& V^\beta(y) \biggl(1-\frac
{r^{(l)}(y)}{r^{(l)}(x_2)} \biggr)\\
&=& e^{\kappa\beta E(y)} \bigl(1-e^{-\tau(E(y)-E(x_2))} \bigr)\\
&=& e^{-(\tau-\kappa\beta) E(y)} \bigl(e^{\tau E(y)}-e^{\tau E(x_2)} \bigr)\\
&\leq& \bigl(e^{\tau(E(x_1)}-e^{\tau E(x_2))} \bigr) e^{-(\tau-\kappa\beta) E(y)}.
\end{eqnarray*}
Similarly, on $r^{(l)}(y)\leq\min(r^{(l)}(x_1),r^{(l)}(x_2))$,
\begin{eqnarray*}\Delta(y)
&\leq&
\big\vert e^{\tau E(x_1)}-e^{\tau
E(x_2)}\big\vert V^\beta(y)r^{(l)}(y)\\
&=&
\big\vert e^{\tau E(x_1)}-e^{\tau E(x_2)}\big\vert e^{-(\tau-\kappa\beta) E(y)}.
\end{eqnarray*}
Putting the three parts together yields the lemma.
%we see that:
%E(x_1)}-e^{\tau E(x_2)}}\int\mu(dy)e^{-(\tau-\kappa) E(y)},\]
%and $e^{-(\tau-\kappa) E(y)}$ is a bounded function since $E$ is
%bounded from below and $\tau>\kappa$.
\end{pf}
\begin{remark}
Lemma \ref{lemunifcont} will be useful in deriving a uniform law of
large numbers for $\{X_n^{(l)}\}$. Actually, this lemma shows that if
the function $E$ is continuous then the kernel $N_\mu^{(l)}$ is a
strong Feller kernel that transforms a bounded function $f$ into a
continuous bounded function $N_\mu^{(l)}$ (uniformly in $\mu$). We
will use this later.
\end{remark}
%
%s4.2 ###
\subsection{Poisson equation}

A straightforward consequence of Section~\ref{DrMinoEE} is that for any $f\in
L^\infty_{V^\beta}$, $\beta\in(0,1]$ the function
%e30 ###
%
\begin{equation}U^{(l)}_{\nu}f(x):=\sum_{k=0}^\infty \bigl[P^{(l)}_{\nu
}-\pi^{(l)}_{\nu} \bigr] ^kf(x)
\end{equation}
is well defined and
%e31 ###
%
\begin{equation}\label{boundUEE}
\abss{U^{(l)}_{\nu}f}_{V^\beta}+\abss
{P^{(l)}_{\nu}U^{(l)}_{\nu}f}_{V^\beta}\leq C\abs{f}_{V^\beta},
\end{equation}
where $C$ is finite and does not depend on $\nu$ nor $f$. $U_\nu
^{(l)}f$ satisfies the (Poisson) equation
%e32 ###
%
\begin{equation}\label{poisson}U_\nu^{(l)}f(x)-P_\nu U_\nu
^{(l)}f(x)=f(x)-\pi^{(l)}_{\nu}(f),\qquad x\in\X.
\end{equation}
%
%For $\eps\in(0,1)$, we introduce the resolvent kernel of $P_
%$U^{(l)}_{\eps,\nu}f(x)$ is well defined and satisfies the resolvent
%equation

Lemmas \ref{DrMinoEE} and \ref{stabEE} implie that for all $\beta\in
(0,1]$, and $\mu,\nu$ probability measures on $(\X,\B)$:
%e33 ###
%
\begin{equation}\label{stabpiEE}\big\Vert\pi^{(l)}_{\mu}-\pi
^{(l)}_{\nu}\big\Vert_{V^\beta}\leq C\Vert\mu-\nu\Vert_{V^\beta};
\end{equation}
for $f\in L^\infty_{V^\beta}$,
%e34 ###
%
\begin{equation}\label{stabuEE}
\abss{U^{(l)}_{\mu}f-U^{(l)}_{\nu
}f}_{V^\beta}\leq C\abs{f}_{V^\beta}\Vert\mu-\nu\Vert_{V^\beta}
\end{equation}
and
%e35 ###
%
\begin{equation}\label{stabUEE}
\abss{P^{(l)}_{\mu}U^{(l)}_{\mu
}f-P^{(l)}_{\nu}U^{(l)}_{\nu}f}_{V^\beta}\leq C\abs{f}_{V^\beta
}\Vert\mu-\nu\Vert_{V^\beta}.
\end{equation}
The inequalities (\ref{stabpiEE}), (\ref{stabuEE}) and (\ref
{stabUEE}) can be derived, for example, by adapting the proofs of
Proposition 3 of \cite{andrieuetal06}. We omit the details. An
important point is the fact that the constant $C$ (whose actual value
can change from one equation to the other) does not depend on $f$ nor
$\nu,\mu$.

%s4.3 ###
\subsection{\texorpdfstring{Proof of Theorem \protect\ref{thm1}}{Proof of Theorem 3.1}}\label{proofthm1}

Let $f\dvtx (\M,\B_\M)\times(\X,\B)\to\Real$ be a measurable
function. We will use the notation $f_\mu(x)$ when evaluating $f$.
%Call $\nu^{(l-1)}_n$ the empirical measure generated by
%$X_{0:n}^{(l-1)}$ and denote $P^{(l)}_{n}$ the kernel $P_{
%$f_n$, $\pi_{n}(f)$ and $U^{(l)}_{n}f(x)$ the quantities $f_{
%random variables by the Polish assumption, the measurability
%assumption on $f$ and by (\ref{stabpiEE}) and (\ref{stabuEE}).
%, (). the Polish assumption the map $x\to\delta_x$ from $(\X,B)\to(\M,
%and from the Polish assumption, $\pi_{x_{0:n-1}}^{(l)}(f)$ is a
%measurable function of $x_{0:n-1}$ and we denote $\pi_{n-1}^{(l)}(f)$
%the random variable $\pi^{(l)}_{X_{0:n-1}^{(l)}}(f)$. Similarly
%$U^{(l)}_{x_{0:n-1}}f(x)$, the fundamental kernel of
%$P_{x_{0:n-1}}^{(l)}$ at $f$ is a measurable function of
%$(x_{0:n-1},x)$ and if $X$ is a $\X$-valued random variable, we call
%$U^{(l)}_{n-1}f(X)$ the random variable
%$U^{(l)}_{X^{(l-1)}_{0:n-1}}f(X)$.
We introduce the partial sum associated to $\{X_n^{(l)}, n\geq0\}$:
\begin{eqnarray*}
S_n^{(l)}(f)&:=&
\sum_{k=1}^nf_{\mu_{k-1}^{(l-1)}}\bigl(X_k^{(l)}\bigr)\\
&&\hspace*{-10.8pt}=\sum
_{k=1}^n\pi^{(l)}_{\mu_{k-1}^{(l-1)}} \bigl(f_{\mu_{k-1}^{(l-1)}} \bigr)\\
&&{}+ \sum
_{k=1}^n \bigl(f_{\mu_{k-1}^{(l-1)}}\bigl(X_k^{(l)}\bigr)-\pi^{(l)}_{\mu
_{k-1}^{(l-1)}} \bigl(f_{\mu_{k-1}^{(l-1)}} \bigr) \bigr).
\end{eqnarray*}
Using the Poisson equation (\ref{poisson}), we have the decomposition
%e36 ###
%
\begin{eqnarray}\label{martapprox}
S_n^{(l)}(f)
&=&\sum_{k=1}^n\pi
^{(l)}_{\mu_{n-1}^{(l-1)}} \bigl(f_{\mu_{k-1}^{(l-1)}} \bigr)+
M_n^{(l)}(f)+R_{n,1}^{(l)}(f)+R_{n,2}^{(l)}(f),
\nonumber\\[-8pt]\\[-8pt]
M_n^{(l)}(f)
&=&\sum_{k=1}^nD_k^{(l)}(f),\nonumber
\end{eqnarray}
where
\begin{eqnarray*}
D_k^{(l)}(f)&=&U^{(l)}_{\mu_{k-1}^{(l-1)}}f_{\mu
_{k-1}^{(l-1)}}\bigl(X_k^{(l)}\bigr)-P^{(l)}_{\mu_{k-1}^{(l-1)}}U_{\mu
_{k-1}^{(l-1)}}^{(l)}f_{\mu_{k-1}^{(l-1)}}\bigl(X_{k-1}^{(l)}\bigr),
\\
R_{n,1}^{(l)}(f)&=&P^{(l)}U_{0}^{(l)}f_0\bigl(X_{0}^{(l)}\bigr)-P^{(l)}_{\mu
_{n}^{(l-1)}}U_{\mu_{n}^{(l-1)}}^{(l)}f_{\mu_{n}^{(l-1)}}\bigl(X_{n}^{(l)}\bigr)
\end{eqnarray*}
and
\[
R_{n,2}^{(l)}(f)=\sum_{k=1}^n P^{(l)}_{\mu_{k}^{(l-1)}}U_{\mu
_{k}^{(l-1)}}^{(l)}f_{\mu_{k}^{(l-1)}}\bigl(X_{k}^{(l)}\bigr)-P^{(l)}_{\mu
_{k-1}^{(l-1)}}U_{\mu_{k-1}^{(l-1)}}^{(l)}f_{\mu_{k-1}^{(l-1)}}\bigl(X_k^{(l)}\bigr).
\]
\begin{lemma}\label{theo2:lem1}
\[
\sup_{1\leq l\leq K}\sup_{k,k'\geq0}\PE
\bigl(V\bigl(X_{k'}^{(l-1)}\bigr)V\bigl(X_k^{(l)}\bigr) \bigr)<\infty.
\]
\end{lemma}
\begin{pf}This is a straightforward consequence of the (uniform in $\nu
$) drift condition on $P^{(l)}_\nu$.
\end{pf}
\begin{lemma}\label{theo2:lem2}Let $p> 1$ such that $p\beta\leq1$.
There exists a finite constant $C$ such that
\[
\E\bigl[\abss{R_{n,2}^{(l)}(f)}^p \bigr]\leq C (\log n )^p.
\]
Moreover $n^{-1}R_{n,2}^{(l)}(f)$ converges $\p$-almost surely to $0$.
\end{lemma}
\begin{pf}
We use (\ref{stabUEE}), (\ref{boundUEE}) and (\ref{eq1keyergothmEE})
to obtain
\begin{eqnarray}\label{eq1lemiiEE}
&&\abss{ P^{(l)}_{\mu
_{k}^{(l-1)}}U_{\mu_{k}^{(l-1)}}
^{(l)}f_{\mu_{k}^{(l-1)}}\bigl(X_{k}^{(l)}\bigr)-P^{(l)}_{\mu
_{k-1}^{(l-1)}}U_{\mu_{k-1}^{(l-1)}}
^{(l)}f_{\mu_{k-1}^{(l-1)}}\bigl(X_k^{(l)}\bigr)}^p\nonumber\\[-8pt]\\[-8pt]
&&\qquad\leq C\sup_{\nu\in\M}\abs{f_\nu}_{V^\beta}^p\big\Vert\mu
^{(l-1)}_{n}-\mu^{(l-1)}_{n-1}\big\Vert
^p_{V^\beta}V^{\beta p}\bigl(X_k^{(l)}\bigr).\nonumber
\end{eqnarray}
But $\mu_n^{(l-1)}=\mu_{n-1}^{(l-1)}+n^{-1} (\delta
_{X_n^{(l-1)}}-\mu_{n-1}^{(l-1)} )$ and we get
\begin{eqnarray*}
\big\Vert\mu_n^{(l-1)}-\mu_{n-1}^{(l-1)}\big\Vert_{V^\beta
}
&=&
\sup_{\abs{f}_{V^\beta}\leq1}\abss{ \bigl(\mu_n^{(l-1)}-\mu
_{n-1}^{(l-1)} \bigr)(f)}\\
%&\leq&\frac{\omega_n^{(l)}}{s_n^{(l)}}\abs{T^{(l)}V^\beta(X_n^{(l-1)})+
&\leq&
\frac{1}{n+1} \Biggl(V^\beta\bigl(X_n^{(l-1)}\bigr)+\frac{1}{n}\sum
_{k=0}^{n-1}V^\beta\bigl(X^{(l-1)}_k\bigr) \Biggr).
\end{eqnarray*}
In view of Lemma \ref{theo2:lem1} and since\vspace*{-1pt} $p\beta\leq1$, $\E
[V^{p\beta}(X_k^{(l)}) (V^\beta(X_n^{(l-1)})+\frac{1}{n}\times\break\sum
_{k=0}^{n-1}V^\beta(X^{(l-1)}_k) )^p ]\leq C$ for some finite constant
$C$ that does not depend on $n$. Therefore, given~(\ref{eq1lemiiEE})
and (\ref{eq1keyergothmEE}), we can use Minkowski's inequality to
conclude the first part of the lemma.

For the second part, by Kronecker's lemma, it is enough to show that
the series
\[
\sum_{k\geq1}k^{-1} \bigl( P^{(l)}_{\mu_{k}^{(l-1)}}U_{\mu
_{k}^{(l-1)}}^{(l)}f_{\mu_{k}^{(l-1)}}\bigl(X_{k}^{(l)}\bigr)-
P^{(l)}_{\mu_{k-1}^{(l-1)}}U_{\mu_{k-1}^{(l-1)}}^{(l)}f_{\mu
_{k-1}^{(l-1)}}\bigl(X_k^{(l)}\bigr) \bigr)
\]
converges almost surely. This will follow if we show that
\[
\sum_{k\geq1}k^{-1}\E\bigl( \big| P^{(l)}_{\mu_{k}^{(l-1)}}U_{\mu
_{k}^{(l-1)}}^{(l)}f_{\mu_{k}^{(l-1)}}\bigl(X_{k}^{(l)}\bigr)-
P^{(l)}_{\mu_{k-1}^{(l-1)}}U_{\mu_{k-1}^{(l-1)}}^{(l)}f_{\mu
_{k-1}^{(l-1)}}\bigl(X_k^{(l)}\bigr) \big| \bigr)
\]
is finite. But from the above calculations, we have seen that
\[
\E\bigl( \big| P^{(l)}_{\mu_{k}^{(l-1)}}U_{\mu_{k}^{(l-1)}}^{(l)}f_{\mu
_{k}^{(l-1)}}\bigl(X_{k}^{(l)}\bigr)-
P^{(l)}_{\mu_{k-1}^{(l-1)}}U_{\mu_{k-1}^{(l-1)}}^{(l)}f_{\mu
_{k-1}^{(l-1)}}\bigl(X_k^{(l)}\bigr) \big| \bigr)\leq Ck^{-1}.
\]
The lemma thus follows.
\end{pf}
\begin{lemma}\label{theo2:lem3}Let $p> 1$ such that $\beta p\leq1$. Then
\[
\sup_n\E\bigl[\abss{R_{n,1}^{(l)}(f)}^p \bigr]<\infty.
\]
Moreover for any $\delta>0$,
\[
\Pr\Bigl[\sup_{m\geq n} \big|m^{-1}R_{m,1}^{(l)}(f) \big|>\delta\Bigr]\to0\qquad \mbox{as } n\to\infty.
\]
\end{lemma}
\begin{pf}
The first part is a direct consequence of~(\ref{eq1keyergothmEE}) and
(\ref{boundUEE}). For the second part, by Markov's inequality, we see that
\begin{eqnarray*}
\Pr\Bigl[\sup_{m\geq n} \big|m^{-1}R_{m,1}^{(l)}(f) \big|>\delta
\Bigr]
&\leq&
\delta^{-p}\E\biggl[\sum_{m\geq n}m^{-p} \big|R_{m,1}^{(l)}(f)\big |^p \biggr]\\
&\leq&
C\delta^{-p}\sum_{m\geq n}m^{-p}\to0\qquad \mbox{as } n\to
\infty.
\end{eqnarray*}
\upqed
\end{pf}
\begin{lemma}\label{theo2:lem4}
Let $p>1$ such that $p\beta\leq1$. There exists a finite constant $C$
such that
\[
\E\bigl[\abss{M_{n}^{(l)}(f)}^p \bigr]\leq Cn^{\max(1,p/2)}.
\]
\end{lemma}
\begin{pf}
By Burkeholder's inequality applied to the martingale $\{M_n^{(l)}(f)\}
$, we get
\[
\E\bigl[ \big|M_n^{(l)}(f) \big|^p \bigr]\leq
C\E\Biggl[ \Biggl(\sum_{k=1}^n \big|D_{k-1}^{(l)}(f)\big |^2
\Biggr)^{p/2} \Biggr].
\]
If $p\geq2$, we apply Minkowski's inequality and use (\ref{boundUEE})
to conclude that
\[
\E\bigl[ \big|M_n^{(l)}(f) \big|^p \bigr]\leq C \Biggl\{\E\Biggl[\sum_{k=1}^n\E^{2/p} \bigl(V^{p\beta
}\bigl(X_{k-1}^{(l)}\bigr) \bigr) \Biggr] \Biggr\}^{p/2}\leq Cn^{p/2}.
\]
If $1<p\leq2$, we use the inequality $(a+b)^\alpha\leq a^\alpha+
b^\alpha$ valid for all $a,b\geq0$, $\alpha\in[0,1]$ to write
\begin{eqnarray*}
\E\bigl[ \big|M_n^{(l)}(f) \big|^p \bigr]&\leq& C\E\Biggl(\sum_{k=1}^n
\big|D_k^{(l)}(f) \big|^p \Biggr)\\
&\leq&C\sum_{k=1}^n\E\bigl(V^{p\beta}\bigl(X_{k-1}^{(l)}\bigr) \bigr)\leq Cn.
\end{eqnarray*}
\upqed
\end{pf}

To deal with the remaining term, we will rely on the following result
which is also of some independent interest.
\begin{lemma}\label{lemlast}Let $\mu,\mu_1,\ldots$ be a sequence of
probability measures on a measurable space $(\X,\B)$ such that $\mu
_n(A)\to\mu(A)$ for all $A\in\B$ and let $f,f_1,\ldots$ be a
sequence of measurable real-valued functions defined on $(\X,\B)$
such that $\sup_n\abs{f_n}_V<\infty$ and $f_n(x)\to f(x)$ for all
$x\in\X$ for some measurable function $V\dvtx (\X,\B)\to(0,\infty)$
such that $\mu(V)<\infty$ and $\sup_n\mu_n(V^\alpha)<\infty$ for
some $\alpha>1$. Then
\[
\lim_{n\to\infty}\mu_n(f_n)=\mu(f).
\]
\end{lemma}
\begin{pf}
By \cite{royden88}, Chapter 11, Proposition 18, we only need to prove
that $\mu_n(V)\to\mu(V)$. By \cite{royden88}, Chapter 11, Proposition
17, we already have $\mu(V)\leq\liminf_{n\to\infty}\mu_n(V)$.
Now we show that ${\lim\sup}_{n\to\infty}\mu_n(V)\leq\mu(V)$ which
will prove the lemma.

Since $V>0$, there exists a sequence of nonnegative simple measurable
functions $\{V_n\}$ that converges increasingly to $V$ $\mu$-a.s. For
$k\geq1$, $N\geq1$, define $E_{k,N}=\{x\in\X\dvtx V(x)-V_p(x)\geq\frac
{1}{k}$, for some $p\geq N\}$. Clearly, $E_{k,N}\in\B$ and
$\mu(E_{k,N})\to0$ as $N\to\infty$ for any $k\geq1$. Fix $k,N\geq
1$. Then for any $n\geq1$ and any $p\geq N$, we have
%e37 ###
%
\begin{eqnarray}\label{eq1lemlast}
\mu_n(V)
&=&\mu_n(V_p)+ \mu
_n(V-V_p)\nonumber\\[-1.5pt]
&=&\mu_n(V_p)+ \int_{E_{k,N}}\mu_n(dx) \bigl(V(x)-V_p(x)
\bigr)\nonumber\\[-1.5pt]
&&{}+\int
_{E_{k,N}^c}\mu_n(dx) \bigl(V(x)-V_p(x) \bigr)\\[-1.5pt]
&\leq&
\mu_n(V_p)+\int_{E_{k,N}}\mu_n(dx)V(x)+\frac{1}{k}\nonumber
\\[-1.5pt]
&\leq&\mu_n(V_p)+C (\mu_n(E_{k,N}) )^q+\frac{1}{k},\nonumber
\end{eqnarray}
with $q=1-1/\alpha$ for some finite constant $C$. The last inequality
uses the inequality of Holder and the assumption that $\sup_n\mu
_n(V^\alpha)<\infty$ for some $\alpha>1$. Since $V_k$ is simple,
$\mu_n(V_k)\to\mu(V_k)$. Also $\mu_n(E_{k,N})\to\mu(E_{k,N})$.
With these and letting $n\to\infty$ and $p\to\infty$ in (\ref
{eq1lemlast}), we have by monotone convergence
\[
{\lim\sup}_{n\to\infty}\mu_n(V)\leq\mu(V)+C (\mu(E_{k,N}) )^q+\frac{1}{k}.
\]
Letting $N\to\infty$ and then $k\to\infty$, we get ${\lim\sup}_{n\to
\infty}\mu_n(V)\leq\mu(V)$.
\end{pf}

\begin{lemma}\label{lemlastlast}
$\pi^{(l)}_{\mu_{n}^{(l-1)}} (f_{\mu_{n}^{(l-1)}} )\to0$ as $n\to
\infty$ with $\p$ probability one.
\end{lemma}
\begin{pf}
To simplify the notations, we write $\pi_n^{(l)}$, $P^{(l)}_n$ and
$f_{n}$ instead of $\pi^{(l)}_{\mu_{n}^{(l-1)}}$, $P^{(l)}_{\mu
_n^{(l-1)}}$ and $f_{\mu_n^{(l-1)}}$ respectively. For $x\in\X$, and
$n,m\geq1$, we have
%e38 ###
%
\begin{eqnarray}\label{llnstareq}
\big|\pi_n^{(l)}(f_n)- \pi^{(l)}\bigl(f_{\pi^{(l-1)}}\bigr) \big|
&\leq&\abss{\pi
_n^{(l)}(f_n)- \bigl(P_n^{(l)} \bigr)^mf_n(x)}\nonumber\\
&&{}+\abss{ \bigl(P_n^{(l)} \bigr)^mf_n(x)-
\bigl(K^{(l)} \bigr)^mf_{\pi^{(l-1)}}(x)}\nonumber\\
&&{}+\abss{ \bigl(K^{(l)} \bigr)^mf_{\pi^{(l-1)}}(x)- \pi^{(l)}\bigl(f_{\pi
^{(l-1)}}\bigr)}\\
&\leq&
2\sup_{\nu\in\M}\abs{f_\nu}_{V^{\beta}}C_\beta V^\beta
(x)\rho_\beta^m\nonumber\\
&&{}+
\abss{ \bigl(P_n^{(l)} \bigr)^mf_n(x)- \bigl(K^{(l)} \bigr)^mf_{\pi^{(l-1)}}(x)},\nonumber
\end{eqnarray}
using (\ref{ratePn}).
We will show next that there exists $\D_0\in\F$, with $\Pr(\D
_0)=1$ such that for each path $\omega\in\D_0$,
%$f_{\nu^{(l-1)}_{n}}(x)(\omega)$ converges to $f_{\pi^{(l-1)}}(x)$ and
%$ (P_n^{(l)} )^m(x,A)(\omega)$ converges to $ (K^{(l)} )^m(x,A)$ as $n
$ (P_n^{(l)} )^mf_n(x)(\omega)$ converges to $ (K^{(l)} )^mf_{\pi
^{(l-1)}}(x)$ as $n\to\infty$ for all $x\in\X$, all $m\geq0$.
Then, going back to (\ref{llnstareq}), we can conclude that for each
$\omega\in\D_0$,
\[
{\lim\sup}_{n\to\infty}\abss{\pi_n^{(l)}(f_n)- \pi^{(l)}\bigl(f_{\pi
^{(l-1)}}\bigr)}(\omega)\leq2C_\beta V^\beta(x)\rho_\beta^m
\]
and the proof will be finished by letting $m\to\infty$.

We can rewrite $P_n^{(l)}(x,A)$ as
\[
P_n^{(l)}(x,A)=\theta_lP^{(l)}(x,A)+(1-\theta
_l)N_n^{(l)}(x,A)+(1-\theta_l)\mathbf{1}_A(x) \bigl(1-N_n^{(l)}(x,\mathbf
{I}) \bigr),
\]
where $N_n^{(l)}(x,A)=\int\mu_n(dy)\mathbf{1}_A(y)\min(1,\frac
{r^{(l)}(y)}{r^{(l)}(x)} )$ and\vspace*{-1pt} $N_n^{(l)}(x,\mathbf{I})=\int\mu
_n(dy)\times\break\min(1,\frac{r^{(l)}(y)}{r^{(l)}(x)} )$.

By the law of large numbers assumed for $\{X_n^{(l-1)}, n\geq0\}$, and
since $(\X,\B)$ is Polish, there exists a dense countable subset $\C
$ in $\X$, a countable generating algebra $\B_0$ of $\B$ and $\D\in
\F$, $\p(\D)=1$ such that for all $x\in\C$ and all $A\in\B_0$:
%e39 ###
%
\begin{eqnarray}\label{eqstar1}
N_n^{(l)}(x,A)&\to& N^{(l)}(x,A)\qquad \mbox
{as } n\to\infty,
\\
\label{eqstar2}
N_n^{(l)}(x,\mathbf{I})&\to&
N^{(l)}(x,\mathbf{I})\qquad \mbox{as } n\to\infty.
\end{eqnarray}

We can also choose $\D$ such that the convergence of $f_{n}(x)(\omega
)$ to $f_{\pi^{(l-1)}}(x)$ for all $x\in\X$ which is assumed in the
theorem hold for all $\omega\in\D$. If we fix a sample path $\omega
\in\D$, and we fix $x\in\C$, the convergence in (\ref{eqstar1})
can actually be extended to all $A\in\B$ by a classical measure
theory argument. Also, again for $\omega\in\D$ and $A\in\B$ fixed,
we can extend the convergence in~(\ref{eqstar1})--(\ref{eqstar2}) to
hold for all $x\in\X$. To see why, take $x\in\X$ arbitrary. Lemma
\ref{lemunifcont} and the continuity of $E$ implies that $N_\mu(x,A)$
is a continuous function of $x$ uniformly in $\mu$. Since $\C$ is
dense, for all $k\geq1$, there is $x_k\in\C$ such that
\[
\abss{N_\mu^{(l)}(x,A)-N_\mu^{(l)}(x_k,A)}\leq\frac{1}{k}
\]
for all $\mu$. In particular, $N_n^{(l)}(x,A)\geq
N_n^{(l)}(x_k,A)-1/k$ for all $n\geq1$. As\break $n\to\infty$, it follows
that $\liminf_{n\to\infty}N_n^{(l)}(x,A)\geq N_{\pi
^{(l-1)}}^{(l)}(x_k,A)-1/k$. As\break $k\to\infty$, by the continuity of
$N_{\pi^{(l-1)}}^{(l)}f(\cdot)$ (Lemma \ref{lemunifcont}), we see
that\break $\liminf_{n\to\infty}N_n^{(l)}(x,A)\geq N_{\pi
^{(l-1)}}^{(l)}(x,A)$. Similarly, we obtain\break ${\lim\sup}_{n\to\infty
}N_n^{(l)}(x,A)\leq N_{\pi^{(l-1)}}^{(l)}(x,A)$. So that $\lim_{n\to
\infty}N_n^{(l)}(x,A)=\break N_{\pi^{(l-1)}}^{(l)}(x,A)$. Similarly, $\lim
_{n\to\infty}N_n^{(l)}(x,I)= N_{\pi^{(l-1)}}^{(l)}(x, I)$.

This shows that for each sample path $\omega\in\D$, $P_n^{(l)}(x,A)$
converges to $K^{(l)}(x,A)$ for all $x\in\X$ all $A\in\B$. By a
successive application of Lemma~\ref{lemlast} (with $V\equiv1$), we
can therefore conclude that for each sample path $\omega\in\D$
%e41 ###
%
\begin{eqnarray}\label{eq1thm4EE}
&&\bigl(P_n^{(l)} \bigr)^m(x,A)\to\bigl(K^{(l)}
\bigr)^m(x,A),\nonumber\\[-8pt]\\[-8pt]
&&\qquad \mbox{as } n\to\infty \mbox{ for all } x\in\X, A\in\B
, m\geq0.\nonumber
\end{eqnarray}

Since $\sup_n{\abs{f_n}_{V^\beta}}<\infty$ ($\beta\in[0,1)$) and
$ (P_\mu^{(l)} )^mV(x)$ is uniformly bounded in $\mu$ and $m$, we can
apply Lemma \ref{lemlast} again to conclude that for each $\omega\in
\D$, $ (P_n^{(l)} )^mf_n(x)$ converges to $ (K^{(l)} )^mf_{\pi
^{(l-1)}}(x)$ for all $x\in\X$, all $m\geq0$, which ends the proof.
\end{pf}

%s4.3.1 ###
\begin{pf*}{Proof of Theorem \protect\ref{thm1}}
We are now in position to prove Theorem \ref{thm1}. Since $\beta\in
[0,1)$, we can take $p=1/\beta$ in Lemmas~\ref{theo2:lem2} and
\ref{theo2:lem3} to conclude that $R_{i,n}^{(l)}(f)/n\to0$, $\p
$-a.s. for $i=1,2$ and by the strong law of\vspace*{-1pt} large numbers for
martingales \cite{chow66}, we conclude that $M_n^{(l)}(f)/n\to0$,
$\p$-a.s. We finish the proof using Lemma \ref{lemlastlast}.
\end{pf*}

%s4.4 ###
\subsection{\texorpdfstring{Proof of Theorem \protect\ref{thm2}}{Proof of Theorem 3.2}}\label{proofthm2}

Take $p=1/\beta>2$ (since $\beta\in[0,1/2)$). By the martingale
approximation (\ref{martapprox}),
\[
S_n^{(l)}(f)-\sum_{k=1}^n\pi^{(l)}_{\mu_{k-1}^{(l-1)}} \bigl(f_{\mu
_{k-1}^{(l-1)}} \bigr)=M_n^{(l)}(f)+R_n^{(l)}(f).
\]
As above, we will simplify the notations by writing $\pi_n^{(l)}(f_n)$
instead of\break $\pi^{(l)}_{\mu_{k-1}^{(l-1)}} (f_{\mu_{k-1}^{(l-1)}} )$
and similarly for $U_n^{(l)}, P_n^{(l)}$, etc.

By Lemmas \ref{theo2:lem2}--\ref{theo2:lem3}, $\E[\abs
{R^{(l)}_n(f)}^p ]=O ((\log(n))^p )$. We then deduce that
$R_n^{(l)}(f)/\sqrt{n}\stackrel{P}{\to}0$ and it remains to show
that a central limit theorem hold for the martingale $\{M_n^{(l)}(f),\F
_n\}$. We need to show that the Lindeberg condition holds:
%e42 ###
%
\begin{equation}\label{lindEE}\frac{1}{n}\sum_{k=1}^n\E
\bigl[\bigl(D_k^{(l)}\bigr)^2(f)\mathbf{1}_{ \{\abs{D_k^{(l)}(f)}>\eps
\sqrt{n} \}} \bigr]\stackrel{P}{\to} 0\qquad \mbox{for all } \eps>0
\mbox{ as } n\to\infty
\end{equation}
and that
%e43 ###
%
\begin{equation}\label{avarEE}
\frac{1}{n}\sum_{k=1}^n\E
\bigl[\bigl(D_k^{(l)}\bigr)^2(f)\vert\F_{k-1} \bigr]\stackrel{P}{\to}\sigma^2(f),
\end{equation}
where $\sigma^2(f)=\pi(f^2)+2\sum_{i=1}^\infty\pi^{(l)}
[f(K^{(l)})^if ]$.
Since $\sup_n\E(\abs{D_n^{(l)}(f)}^p )<\infty$ for $p>2$, it
follows that the Lindeberg
condition (\ref{lindEE}) holds.

For the law of large numbers, we need some notations. Let $U^{(l)}$
denote the fundamental kernel of the limiting kernel $K^{(l)}$ and
define the functions $\Delta_n^{(1)}(x)=P_n^{(l)} (U_n^{(l)} )^2f(x)$
and $\Delta_n^{(2)}(x)= [P_n^{(l)}U_n^{(l)}f(x) ]^2$. Simularly,
define\vspace*{1pt} $\Delta^{(1)}(x)=K^{(l)} (U^{(l)} )^2f(x)$ and $\Delta
^{(2)}(x)= [K^{(l)}U^{(l)}f(x) ]^2$. Then we can rewrite
\begin{eqnarray*}
&&\frac{1}{n}\sum_{k=1}^{n}\E\bigl(\bigl(D_k^{(l)}\bigr)^2(f)\vert
\F_{k-1} \bigr)\\
&&\qquad =\frac{1}{n}\sum_{k=1}^nP_{k-1}^{(l)} \bigl(U_{k-1}^{(l)}
\bigr)^2f\bigl(X_{k-1}^{(l)}\bigr)- \bigl[P_{k-1}^{(l)}U_{k-1}^{(l)}f\bigl(X_{k-1}^{(l)}\bigr) \bigr]^2\\
&&\qquad =
\frac{1}{n}\sum_{k=1}^n\Delta_{k-1}^{(1)}\bigl(X_{k-1}^{(l)}\bigr)+\Delta
_{k-1}^{(2)}\bigl(X_{k-1}^{(l)}\bigr).
\end{eqnarray*}
Fix $f\in L^\infty_{V^\beta}$. We have seen in the proof of Theorem
\ref{thm1} that $\pi^{(l)}_n(f)$ converges almost surely to $\pi
^{(l)}(f)$. Combined with~(\ref{eq1thm4EE}) and using dominated
convergence it follows that there is $\D\in\F$, $\Pr(\D)=1$ such
that for all sample path $\omega\in\D$, $U_{n}^{(l)}f(x)$ converges
to $U^{(l)}f(x)$ for all $x\in\X$. By virtue of Lemma \ref{lemlast},
it follows that for all $\omega\in\D$, $\Delta_n^{(j)}(x)$
converges to $\Delta^{(j)}(x)$ for all $x\in\X$, $j=1,2$. Then the
strong law of large numbers (Theorem \ref{thm1}), implies that $\frac
{1}{n}\sum_{k=1}^n\E((D_k^{(l)})^2(f)\vert\F_{k-1} )$ converges
almost surely to $\pi^{(l)} (K^{(l)} (U^{(l)} )^2f- [K^{(l)}U^{(l)}f
]^2 )$ which is equal to $\sigma^2(f)=\pi^{(l)}(f^2)+2\sum
_{i=1}^\infty\pi^{(l)} [f(K^{(l)})^if ]$.

%s4.5 ###
\subsection{\texorpdfstring{Proof of Theorem \protect\ref{thm3}}{Proof of Theorem 3.3}}\label{proofthm3}

We continue with the notations of Section \ref{secthm3}.
\begin{lemma}\label{lemHolderGam} Under the assumptions of Theorem
\ref{thm3}, there exists a finite constant $c_0$ such that
\[
|\Gamma(x_1,x)-\Gamma(x,x)|\leq c_0|x_1-x|\qquad \mbox{for all }
x,x_1\in\X.
\]
\end{lemma}
\begin{pf}
Given the expression of $\Gamma$ in (\ref{Gamma}), it is enough to
show that $|U_x^{(0)}(y)-U_{x_1}^{(0)}(y)|\leq c_0|x-x_1|$. But since
\[
\big|U_x^{(0)}(y)-U_{x_1}^{(0)}(y)\big |= \bigg|\sum_{j\geq0} \bigl[\bar P^{(0)} \bigr]^j
\bigl(H_x(y)-H_{x_1}(y) \bigr) \bigg|\leq C |H_x-H_{x_1} |_\infty
\]
(where for a kernel $P$ with invariant distribution $\pi$, $\bar
P=P-\pi$), the lemma follows if we show that there exists a finite
constant $c_0$ such that for any $x_1,x_2,y\in\X$,
\[
|H_{x_1}(y)-H_{x_2}(y) |\leq c_0|x_1-x_2|.
\]
It is easy to check as in Lemma \ref{lemunifcont} that for any
$x_1,x_2,y\in\X$,
%$T^{(1)}(y,x_1,U)-T^{(1)}(y,x_2,U)= (\alpha(x_1,y)-\alpha(x_2,y)
%)U(y)+ U(x_1)(1-\alpha(x_1,y))- U(x_2)(1-\alpha(x_2,y))$, where $
%Lemma \ref{lemunifcont}. On $r_1(y)\leq r(x_1)\wedge r(x_2)$,
%$|T^{(1)}(y,x_1,U)-T^{(1)}(y,x_2,U)|\leq|U(x_1)-U(x_2)|+ 2|U|_\infty
%e^{-\tau E(y)}|e^{\tau E(x_1)}-e^{\tau E(x_1)}|$, where $
%r_1(x_2)$, or $r_1(x_2)\leq r_1(y)\leq r_1(x_2)$,
%$|T^{(1)}(y,x_1,U)-T^{(1)}(y,x_2,U)|\leq2|U|_\infty e^{-\tau E(y)}|e^{
%r_1(x_2)$, $T^{(1)}(y,x_1,U)-T^{(1)}(y,x_2,U)=0$.
%Since $H_{x_1}(y)-H_{x_2}(y)=T^{(1)}(y,x_1,U)-T^{(1)}(y,x_2,U) -\int
%
\begin{eqnarray*}
|H_{x_1}(y)-H_{x_2}(y) |&\leq&2|U(x_1)-U(x_2)|\\
&& {}+ |U|_\infty\biggl(e^{-\tau
E(y)}+\int e^{-\tau E(y)}\pi^{(0)}(dy) \biggr)\big|e^{\tau E(x_1)}-e^{\tau E(x_2)}\big|.
\end{eqnarray*}
Now the result follow from (\ref{LipAssump}), the Lipschitz assumption
on $E$ and the compactness of $\X$.
\end{pf}

\begin{proposition} Under the assumptions of Theorem \ref{thm3}, $\eta
_n$ converges weakly in $\C(\X,\Real)$ to a mean zero Gaussian
process $G$ with covariance function~$\Gamma$ and sample paths in $\C
(\X,\Real)$ and
%e44 ###
%
\begin{equation}\label{boundG}
\E\Bigl(\sup_{x\in\X} |G(x) | \Bigr)<\infty.
\end{equation}
\end{proposition}
\begin{pf}
The existence of $G$ and the bound (\ref{boundG}) follows from Lemma
\ref{lemHolderGam} and Dudley's Theorem on the existence of Gaussian
processes with continuous sample paths (see, e.g., \cite{marcusetrosen}, Theorem 6.1.2). Indeed, if $d_\Gamma(x,y):=(\Gamma
(x,x)+\Gamma(y,y)-2\Gamma(x,y))^{1/2}$ denotes the pseudo-metric
associated to $\Gamma$, Lemma \ref{lemHolderGam} implies that
$d_\Gamma(x,y)\leq\sqrt{2 c_0}|x-y|^{1/2}$ and since $\X$ is
compact, this in turn implies that $\mathcaligr{N}(\X,d_\Gamma
,\epsilon)\leq (K\epsilon^{-1} )^{d/2}$ for some finite constant
$K$, where $\mathcaligr{N}(\X,d_\Gamma,\cdot)$ is the metric
entropy of $\X$ under $d_\Gamma$.

We now show that $\eta_n$ converges weakly in $\C(\X,\Real)$ to a
mean zero Gaussian process with continuous sample path and covariance
function $\Gamma$. Indeed, the convergence of the finite-dimensional
distribution is given by the standard central limit for uniformly
ergodic Markov chains. We use a moment criterion to check that the
family $\{\eta_n, n\geq0\}$ is tight (\cite{kallenberg}, Corollary
16.9). It suffices to check that:
\begin{longlist}
\item[(i)] For some $x_0\in\X$, $\{\eta_n(x_0), n\geq0\}$ is tight.
\item[(ii)] For some positive finite constant $a,b, c_0$,
\[
\E[ |\eta_n(x_1)-\eta_n(x_2) |^a ]\leq c_0|x_1-x_2|^{d+b}\qquad \mbox{for all } x_1,x_2\in\X, n\geq0 .
\]
\end{longlist}
The condition (i) is trivially true. To check (ii), we use the
resolvent $U^{(0)}_x$ to write $H_{x_1}(y)-H_{x_2}(y)=
(U^{(0)}_{x_1}(y)-U^{(0)}_{x_2}(y) )-
(P^{(0)}U^{(0)}_{x_1}(y)-P^{(0)}U^{(0)}_{x_2}(y) )$. It follows that
\[
\eta_n(x_1)-\eta_n(x_2)=M_n(x_1,x_2)+\epsilon_n(x_1,x_2),
\]
where $M_n(x_1,x_2)=\sum_{k=1}^n
(U^{(0)}_{x_1}(X_k^{(0)})-U^{(0)}_{x_2}(X_k^{(0)}) )-
(P^{(0)}U^{(0)}_{x_1}(X_{k-1}^{(0)}) -\break P^{(0)}U^{(0)}_{x_2}(X_{k-1}^{(0)})
)$ and $\epsilon
_n(x_1,x_2)=P^{(0)}U^{(0)}_{x_1}(X_{0}^{(0)})-P^{(0)}U^{(0)}_{x_2}(X_{0}^{(0)})-\break
P^{(0)}U^{(0)}_{x_1}(X_{n}^{(0)})-P^{(0)}U^{(0)}_{x_2}(X_{n}^{(0)})$.

The term $M_n(x_1,x_2)$ is a martingale and $\epsilon_n(x_1,x_2)$ is
bounded in $n$ by a constant. By Burkholder's inequality and some
additional straightforward arguments it follows that for any $a\geq2$
\[
\\E [ |\eta_n(x_1)-\eta_n(x_2) |^a ]\leq
C\big|U_{x_1}^{(0)}-U_{x_2}^{(0)}\big|_\infty^a\leq C|x_1-x_2|^{a}.
\]
Then it suffices to take $a>d$.
\end{pf}

We will also need the following simple result.
\begin{lemma}\label{lemtech1}If $\{x_k\}$ is a sequence of real
numbers such that $x_n\to0$ as $n\to\infty$ then $n^{-1/2}\sum
_{k=1}^nk^{-1/2}x_k\to0$ as $\to\infty$.
\end{lemma}
\begin{pf}
Take $\eps>0$. Let $n_0\geq1$ s.t. $n\geq n_0$ implies $|x_n|\leq
\eps$. Then for $n\geq n_0$, $n^{-1/2}|\sum_{k=1}^nk^{-1/2}x_k|\leq
n^{-1/2}\sum_{k=1}^{n_0}k^{-1/2}|x_k|+n^{-1/2}\sum
_{k=n_0+1}^nk^{-1/2}\eps\leq n^{-1/2}\sum_{k=1}^{n_0}k^{-1/2}|x_k| +
2\eps$. Letting $n\to\infty$ and $\eps\to0$ yields the result.
\end{pf}

\begin{pf*}{Proof of Theorem \protect\ref{thm3}}
For the rest of the proof, let $G$ be a mean zero Gaussian process on
$\X$ with covariance function $\Gamma$ and almost surely continuous
sample paths. We take $G$ independent from the process $\{
(X_n^{(0)},X_n^{(1)}), n\geq0\}$. From the Gaussian process $G$, we
define $\pi(G):=\int G(x)\pi^{(1)}(dx)$ as follows. For each sample
path $\omega\in\Omega$, if $G_\omega(\cdot)$ is continuous then
$\pi(G)(\omega)=\int\pi^{(1)}(dx)G_\omega(x)$. Otherwise, we set
$\pi(G)(\omega)=0$. Since $f\to\pi^{(1)}(f)$ is a continuous map
from $\C(\X,\Real)\to\Real$, $\pi^{(1)}(G)$ is a well-defined
random variable.

Back to the partial sum $S_n$, we have seen that
\[
S_n=M_n +(1-\theta_1)\sum_{k=1}^nk^{-1/2}\eta_k\bigl(X_k^{(1)}\bigr) +\epsilon
_n^{(1)},
\]
where $M_n:=\sum_{k=1}^nU(X_k^{(1)})-P_{\mu
_{k-1}^{(0)}}U(X_{k-1}^{(1)})$ and $\epsilon_n^{(1)}= (P_{\mu
_{0}^{(0)}}U(X_0^{(1)})-P_{\mu_{n}^{(0)}}\times\break U(X_{n}^{(1)}) )$.
Clearly
\[
\sup_{n\geq1} \big| \bigl(P_{\mu_{0}^{(0)}}U\bigl(X_0^{(1)}\bigr)-P_{\mu
_{n}^{(0)}}U\bigl(X_{n}^{(1)}\bigr) \bigr) \big|\leq C,
\]
thus the term $\epsilon_n^{(1)}$ is negligible. That is,
\begin{eqnarray*}
S_n
&=& M_n +(1-\theta_1)\sum_{k=1}^nk^{-1/2}\eta_k\bigl(X_k^{(1)}\bigr)
+o_P\bigl(\sqrt{n}\bigr),\\
&=& M_n
+(1-\theta_1)\sum_{k=1}^n\frac{1}{\sqrt{k}}G\bigl(X_k^{(1)}\bigr)\\
&&{}+(1-\theta_1)\sum_{k=1}^nk^{-1/2} \bigl(\eta_k\bigl(X_k^{(1)}\bigr) -G\bigl(X_k^{(1)}\bigr) \bigr)
+ o_P\bigl(\sqrt{n}\bigr).
\end{eqnarray*}
In the above, we denote $o_P(n^r)$ any random variable $X_n$ such that
$n^{-r}X_n$ converges in probability to zero. To deal with the term
$\sum_{k=1}^nk^{-1/2} (\eta_n(X_k^{(1)})-G(X_k^{(1)}) )$, we use the
Skorohod representation of weak convergence. First note that
\begin{eqnarray*}
&&\Bigg|n^{-1/2}\sum_{k=1}^nk^{-1/2} \bigl(\eta_n\bigl(X_k^{(1)}\bigr)
-G\bigl(X_k^{(1)}\bigr) \bigr) \Bigg|\\
&&\qquad\leq
n^{-1/2}\sum_{k=1}^nk^{-1/2}\sup_{x\in\X} |\eta_n(x)-G(x) |.
\end{eqnarray*}

By the Skorohod representation theorem, there exists a version $\tilde
G$\break of $G$ and a version $\{\tilde\eta_n, n\geq0\}$ of the random
process $\{\eta_n, n\geq0\}$\break such that $\sup_{x\in\X} |\tilde\eta
_n(x)-\tilde G(x) |\to0$ a.s. Therefore, by Lemma \ref{lemtech1},\break
$n^{-1/2}\sum_{k=1}^nk^{-1/2}\sup_{x\in\X} |\tilde\eta
_n(x)-\tilde G(x) |$ converges almost surely and thus in probability to
zero. It follows that $n^{-1/2}\sum_{k=1}^nk^{-1/2} (\eta
_n(X_k^{(1)})-G(X_k^{(1)}) )$ converges also in probability to zero. We
thus arrive at
\[
S_n=M_n+(1-\theta_1)\sum_{k=1}^n\frac{1}{\sqrt{k}}G\bigl(X_k^{(1)}\bigr) +
o_P \bigl(\sqrt{n} \bigr).
\]

To deal with the term $\sum_{k=1}^n\frac{1}{\sqrt{k}}G(X_k^{(1)})$,
we introduce $V_0=0$ and $V_k=\sum_{j=1}^k
(G(X_j^{(1)})-\pi^{(1)}(G))$:
\begin{eqnarray*}
&&\sum_{k=1}^n\frac{1}{\sqrt{k}} \bigl(G\bigl(X_k^{(1)}\bigr)-\pi
^{(1)}(G) \bigr)\\
&&\qquad=
\sum_{k=1}^n\frac{1}{\sqrt{k}} (V_k-V_{k-1} )\\
&&\qquad=
\sum_{k=1}^n\frac{1}{\sqrt{k}}V_k-\sum_{k=2}^n \biggl(\frac{1}{\sqrt
{k}}-\frac{1}{\sqrt{k-1}} \biggr)V_{k-1}-
\sum_{k=2}^n\frac{1}{\sqrt{k-1}}V_{k-1}\\
&&\qquad=
\frac{1}{\sqrt{n}}V_n+\sum_{k=2}^n\frac{1}{\sqrt{k(k-1)} (\sqrt
{k}+\sqrt{k-1} )}V_{k-1}\\
&&\qquad=
\frac{1}{\sqrt{n}}V_n+\sum_{k=2}^n \biggl(\frac{1}{\sqrt{k} (1+\sqrt
{1+1/(k-1)} )} \biggr)\frac{1}{k-1}V_{k-1}.
\end{eqnarray*}
We deduce that
\begin{eqnarray*}
n^{-1/2}S_n
&=&
n^{-1/2}M_n+(1-\theta_1)\pi^{(1)}(G)n^{-1/2}\sum
_{k=1}^nk^{-1/2}+n^{-1}V_n \\
&&{}+\frac{1}{\sqrt{n}}\sum_{k=2}^n \biggl(\frac
{1}{\sqrt{k} (1+\sqrt{1+1/(k-1)} )} \biggr)\frac{1}{k-1}V_{k-1}
+o_P (1 ).
\end{eqnarray*}
For almost every path $\omega\in\Omega$, $G_\omega(\cdot)$ is a
continuous function from $\X\to\Real$. Therefore, by the
independence assumption and the law of large numbers of Theorem~\ref
{thm1}, $n^{-1}\sum_{j=1}^k G(X_j^{(1)})-\pi^{(1)}(G)$ converges in
$L^1$ to zero. Using Lemma~\ref{lemtech1} again, we conclude that
$\frac{1}{\sqrt{n}}\sum_{k=2}^n (\frac{1}{\sqrt{k} (1+\sqrt
{1+1/(k-1)} )} )\frac{1}{k-1}V_{k-1}$ converges also in $L^1$
to zero. The term $\frac{1}{\sqrt{n}}\sum_{k=1}^nk^{-1/2}$ converges
to $2$. We thus arrive at
\[
n^{-1/2} S_n=n^{-1/2}M_n+2(1-\theta_1)\pi^{(1)}(G) +o_P(1).
\]
Proceeding as in the proof of Theorem \ref{thm2}, we see that $\frac
{1}{\sqrt{n}}M_n$ converges weakly to $Z$, where $Z\sim N(0,\sigma
^2_\star(f))$ and is independent from $G$. We thus conclude that
$n^{-1/2}S_n$ converges weakly to $Z+2(1-\theta_1)\int\pi
^{(1)}(dx)G(x)$, where $Z$ and $\int\pi^{(1)}(dx)G(x)$ are independent.

Since $f\to\pi^{(1)}(f)$ is a continuous bounded function from $\C
(\X,\Real)\to\Real$, it follows from the above that $\pi
^{(1)}(\eta_n)$ converges weakly to $\pi^{(1)}(G)$. But $\pi
^{(1)}(\eta_n)=n^{-1/2}\sum_{k=1}^n\int\pi^{(1)}(dx)
H_x(X_k^{(0)})$. By the central limit theorem for the uniformly ergodic
chain $\{X_n^{(0)}, n\geq0\}$, the latter term\break $n^{-1/2}\sum
_{k=1}^n\int\pi^{(1)}(dx) H_x(X_k^{(0)})$ converges weakly to
$N(0,\Gamma(\bar g,\bar g))$, where\break $\bar g(\cdot)=\int\pi
^{(1)}(dx) H_x(\cdot)$ and we are finished.
\end{pf*}

%s4.6 ###
\subsection{\texorpdfstring{Proof of Proposition \protect\ref{propAsympVar}}{Proof of Proposition 3.1}}\label
{proofpropAsympVar}

In the present case, one can check that $U(x)=\sum_{j\geq0}(P_{\pi
^{(0)}}^{(1)})^jf(x)=\sum_{j\geq0}\theta_1^jP^j f(x)$ and
$H_x(y)=U(y)$. Then the resolvent function $U^{(0)}_x$ becomes
$U_x^{(0)}(y)=U^{(0)}(y)=\sum_{j\geq0} P^j U(y)$\break which allows use to
write $\sum_{j=1}^k H_x(X_j^{(0)})=M_k^{(0)}+\epsilon_k^{(0)}$, where
$M_k^{(0)}=\break\sum_{j=1}^kU^{(0)}(X_k^{(0)})-P U^{(0)}(X_{k-1}^{(0)})$
and $\epsilon_k^{(0)}=P U^{(0)}(X_{0}^{(0)})-P U^{(0)}(X_{k}^{(0)})$.
Thus we have
\[
S_n=M_n+(1-\theta_1)\sum_{k=1}^nk^{-1}M_k^{(0)} +\epsilon_n,
\]
where $\epsilon_n=\epsilon^{(1)}_n+\sum_{k=1}^nk^{-1}\epsilon
_k^{(0)}$. The term $\epsilon_n$ is negligible and is suffices to
study the limit of
\begin{eqnarray*}
\E\Biggl[ \Biggl(M_n+(1-\theta_1)\sum_{k=1}^nk^{-1}M_k^{(0)} \Biggr)^2 \Biggr]
&=&\E(M_n^2 )
+(1-\theta_1)^2\E\Biggl[ \Biggl(\sum_{k=1}^nk^{-1}M_k^{(0)} \Biggr)^2 \Biggr] \\
&&{} + 2(1-\theta
_1)\E\Biggl[M_n\sum_{k=1}^nk^{-1}M_k^{(0)} \Biggr].
\end{eqnarray*}
Define $D^{(0)}(x,y)=U^{(0)}(y)-P U^{(0)}(x)$ and $D^{(1)}(x,y)=U(y)-P
U(x)$. It is easy to see that for any $i,j\geq1$, $\E
(D^{(0)}(X_{i-1}^{(0)},X_i^{(0)})D^{(1)}(X_{j-1}^{(1)},X_j^{(1)}) )=0$.
From which we deduce that $\E[M_n\sum_{k=1}^nk^{-1}M_k^{(0)} ]=0$.

We write $\sum_{k=1}^nk^{-1}M_k^{(0)}=\sum_{j=1}^n\sum
_{k=j}^nk^{-1}D^{(0)}(X_{j-1}^{(0)},X_j^{(0)})$ and since the terms
$D^{(0)}(X_{j-1}^{(0)},X_j^{(0)})$ are martingale differences, we get
\begin{eqnarray*}
&&\E\Biggl[ \Biggl(\sum_{k=1}^nk^{-1}M_k^{(0)} \Biggr)^2 \Biggr]\\
&&\qquad=
\E\Biggl[ \Biggl(\sum_{j=1}^n \Biggl(\sum_{k=j}^nk^{-1}
\Biggr)D^{(0)}\bigl(X_{j-1}^{(0)},X_j^{(0)}\bigr) \Biggr)^2 \Biggr]\\
&&\qquad=\sum_{j=1}^n \Biggl(\sum_{k=j}^nk^{-1} \Biggr)^2\E\bigl[
\bigl(D^{(0)}\bigl(X_{j-1}^{(0)},X_j^{(0)}\bigr) \bigr)^2 \bigr]\\
&&\qquad=\int\pi(dx)\int P(x,dy)\bigl(D^{(0)}(x,y)\bigr)^2\sum_{j=1}^n \Biggl(\sum
_{k=j}^nk^{-1} \Biggr)^2 \\
&&\qquad\quad{}+ \sum_{j=1}^n \Biggl(\sum_{k=j}^nk^{-1} \Biggr)^2 \biggl(\E\bigl[
\bigl(D^{(0)}\bigl(X_{j-1}^{(0)},X_j^{(0)}\bigr) \bigr)^2 \bigr]\\
&&\qquad\quad{}\hspace*{84pt}-\int\pi(dx)\int
P(x,dy)\bigl(D^{(0)}(x,y)\bigr)^2 \biggr).
\end{eqnarray*}

Since $D^{(0)}$ is a bounded continuous function and $\{X_n^{(0)}\}$ is
uniformly ergodic, the second term on the r.h.s. divided by $n$
converges to zero. Then we notice that $\lim_{n\to\infty}n^{-1}\sum
_{i=1}^n (\sum_{k=j}^nk^{-1} )^2=2$ and we conclude that
\[
\lim_{n\to\infty}\E(n^{-1}S_n^2 )
=\int\pi(dx)\int P(x,dy) \bigl\{
\bigl(D^{(1)}(x,y)\bigr)^2+2(1-\theta_1)^2\bigl(D^{(0)}(x,y)\bigr)^2 \bigr\}.
\]

\section*{Acknowledgments}
The author is grateful to Eric Moulines and Gersende Fort for helpful
discussions and
to an anonymous referee for helping improve the quality of this work.

% imsref loaded by smiklovaite, 2009-10-07 14:47:31
%

\printaddresses

\end{document}